\documentclass[twocolumn,showpacs]{revtex4}
\usepackage{graphics}

\begin{document}

\title{Photon-number distributions of twin beams generated in
spontaneous parametric down-conversion and measured by an
intensified CCD camera}

\author{Jan Pe\v{r}ina Jr.\thanks{e-mail:
perinaj@prfnw.upol.cz}} \affiliation{Institute of Physics of Academy of Sciences of the Czech Republic,
Joint Laboratory of Optics of Palack\'{y} University and Institute of Physics of Academy of
Science of the Czech Republic, 17.listopadu 12, 772 07 Olomouc, Czech Republic}
\author{Ond\v{r}ej Haderka}
\affiliation{RCPTM, Joint Laboratory of Optics of Palack\'{y}
University and Institute of Physics of Academy of Science of the
Czech Republic, Faculty of Science, Palack\'{y} University, 17.
listopadu 12, 77146 Olomouc, Czech Republic}
\author{Martin Hamar}
\affiliation{Institute of Physics of Academy of Sciences of the Czech Republic,
Joint Laboratory of Optics of Palack\'{y} University and Institute of Physics of Academy of Science of the Czech Republic, 17.listopadu 12, 772 07 Olomouc, Czech Republic}
\author{V\'{a}clav Mich\'{a}lek}
\affiliation{Institute of Physics of Academy of Sciences of the Czech Republic,
Joint Laboratory of Optics of Palack\'{y} University and Institute of Physics of Academy of Science of the Czech Republic, 17.listopadu 12, 772 07 Olomouc, Czech Republic}

\begin{abstract}
The measurement of photon-number statistics of fields composed of
photon pairs, generated in spontaneous parametric down-conversion
and detected by an intensified CCD camera is described. Final
quantum detection efficiencies, electronic noises, finite numbers
of detector pixels, transverse intensity spatial profiles of the
detected beams as well as losses of single photons from a pair are
taken into account in a developed general theory of photon-number
detection. The measured data provided by an iCCD camera with
single-photon detection sensitivity are analyzed along the
developed theory. Joint signal-idler photon-number distributions
are recovered using the reconstruction method based on the
principle of maximum likelihood. The range of applicability of the
method is discussed. The reconstructed joint signal-idler
photon-number distribution is compared with that obtained by a
method that uses superposition of signal and noise and minimizes
photoelectron entropy. Statistics of the reconstructed fields are
identified to be multi-mode Gaussian. Elements of the measured as
well as the reconstructed joint signal-idler photon-number
distributions violate classical inequalities. Sub-shot-noise
correlations in the difference of the signal and idler photon
numbers as well as partial suppression of odd elements in the
distribution of the sum of signal and idler photon numbers are
observed.
\end{abstract}



\maketitle

\section{Introduction}

Light generated in the process of spontaneous parametric
down-conversion (SPDC) is emitted in photon pairs
\cite{Mandel1995}. Photons comprising one photon pair are strongly
quantum correlated (entangled). Entanglement of photons in a pair
has been used in many experiments that have provided a deep
insight into the laws of quantum mechanics
\cite{Milburn1995,Perina1994}. Among others, the measured
violation of Bell's inequalities ruled out neoclassical local
hidden-variables theories. Photon pairs have also found their way
to practical applications, e.g., in quantum cryptography
\cite{Bouwmeester1997}, measurement of ultrashort time delays, or
absolute measurements of detection quantum efficiencies
\cite{Migdall1999,Brida2010}. These experiments utilize photon
fields that contain only one photon pair in a measured time window
with a high probability.

There have been experiments (teleportation, measurement of GHZ
correlations, etc.) measuring triple and quadruple coincidence
counts caused by fields containing two photon pairs in a time
window given by an ultrashort pump pulse. However, states used for
such experiments contain a very low fraction of states with two
photon pairs in comparison with the fraction belonging to the
state with one photon pair and the vacuum state. The reason is to
eliminate the influence of three and more-than-three photon-pair
states to the considered experimental setups. Measurements done in
such setups have to be conditional and they require long
data-acquisition times.

The use of more powerful pump pulses as well as development of
materials with higher values of $ \chi^{(2)} $ susceptibilities
have opened the way to generate fields containing many photon
pairs originating in one pump pulse. For such fields, a joint
signal-idler photon-number distribution is the main characteristic
that determines the experimental results utilizing these fields.
Determination of photon-pair statistics is important also for weak
cw fields provided that they are detected in long-time detection
windows \cite{Larchuk1995}. In this case photon-pair statistics
have been identified to be Poissonian after eliminating dead-time
detection effects \cite{Larchuk1995}.

Returning back to more intense fields, recent experiments
\cite{Paleari2004,Jedrkiewicz2004,Agliati2005,Haderka2005,Haderka2005a,Bondani2007,Blanchet2008,Brida2009a}
(and references therein) are even able to provide experimental
joint signal-idler photoelectron distributions of twin beams
containing up to several thousands of photon pairs. As for
detectors, weaker fields containing up to ten photons can be
measured by special single-photon avalanche detectors (VLPC)
\cite{Kim1999}, hybrid photo-multipliers
\cite{Ramilli2010,Allevi2010}, super-conducting bolometers
\cite{Miller2003} or time-multiplexed fiber-optics detection loops
\cite{Haderka2004,Rehacek2003,Achilles2003,Achilles2004,Fitch2003,Fitch2004}.
Intensified CCD cameras \cite{Jost1998,Haderka2005,Hamar2010} can
in principle capture states with hundreds of photons.
Ultra-sensitive photodiodes with their linear response and very
low level of noise are suitable for the detection of states with
hundreds or even better thousands of photons. A special method
utilizing precisely attenuated beams has also been suggested and
developed \cite{Zambra2005,Zambra2006}. It allows to resolve
photon numbers even in the measurement based on single-photon
sensitive avalanche photodiodes. We note that also the well-known
homodyne detection has been found useful in the determination of
intensity correlations of twin beams
\cite{Vasilyev2000,Zhang2002}.

All these approaches give experimental photoelectron distributions
obtained by detectors with finite quantum detection efficiencies.
While silicon PIN photodiodes or back-illuminated CCD cameras can
offer detection efficiencies close to unity, their internal noise
prevents their use in the single-photon regime. On the other hand,
detectors with large internal gain, like iCCD cameras, EMCCD
cameras or avalanche photo-detectors allow single-photon
sensitivity or even photon-number resolution by effectively
decreasing their noise. A price for this sensitivity is paid,
however, in the form of lower quantum efficiencies. ICCD cameras
are a good trade-off in this respect. Their level of noise is low,
but not negligible. On the other hand, quantum detection
efficiencies around 20~\% are sufficient enough to profit from
their low level of noise.

Once we know quantum detection efficiency and the level of noise,
we can reconstruct the field in front of a detector. The usual and
physically-motivated approach is based on the assumption of the
character of the reconstructed field. Working with photon pairs,
we can naturally assume that the reconstructed field is composed
of certain number of independent modes containing photon pairs and
small additional noise in the form of single photons. Using this
picture, a multi-mode theory of signal and noise tailored
specially for paired fields can be applied (see
\cite{Perina2005,Perina2006a} for the spontaneous process and
\cite{Perina2006,Perina2008,Perina2008a} for the stimulated
process). It can be accompanied with the principle of minimum
entropy to get the reconstructed field. As an alternative, one may
rely on a mathematically based method that uses the
maximum-likelihood principle. In the framework of this method, the
reconstructed field is reached as a steady point accessible by an
iteration procedure. As a final step in characterization of the
fields, joint signal-idler quasi-distributions of integrated
intensities may be reached using the reconstructed joint
signal-idler photon-number distributions
\cite{Perina1991,Haderka2005a}.

Here, we pay attention to the determination of a joint
signal-idler photon-number distribution beyond a nonlinear crystal
using an iCCD camera as a tool resolving photon numbers. The
method of maximum likelihood is applied. It allows to deal with
even more difficult experimental conditions like those reached
when more than one photon can be registered in a single pixel.
Transformation matrices describing details of the detection
process and being an important ingredient of the iteration
procedure are derived under several conditions. The reconstructed
fields are compared with those obtained by the method of
superposition of signal and noise.

The paper is organized as follows. Sec.~II contains a general
model describing a photon-number-resolving detection device.
Sec.~III is devoted to the description of photon-number-resolving
detection by an iCCD camera under real experimental conditions.
The role of inhomogeneous transverse profiles of the detected
fields is discussed in Sec.~IV. In Sec.~V, the iteration procedure
of the maximum-likelihood method is explained and used to recover
joint signal-idler photon-number distributions. Subsec.~VA is
devoted to nonclassical characteristics of the emitted fields.
Statistics of the fields are discussed in Subsec.~VB in which also
the problem of reconstruction of more intense fields is addressed.
Comparison of the reconstructed fields obtained by the
maximum-likelihood method and the method of superposition of
signal and noise is provided in Sec.~VI. Sec.~VII brings
conclusions. The formula for an effective quantum detection
efficiency is derived in Appendix A.

\section{Probabilities of multi-photon-coincidence counts in an array
of single-photon detectors}

\begin{figure}         
 \resizebox{0.98\hsize}{!}{\includegraphics{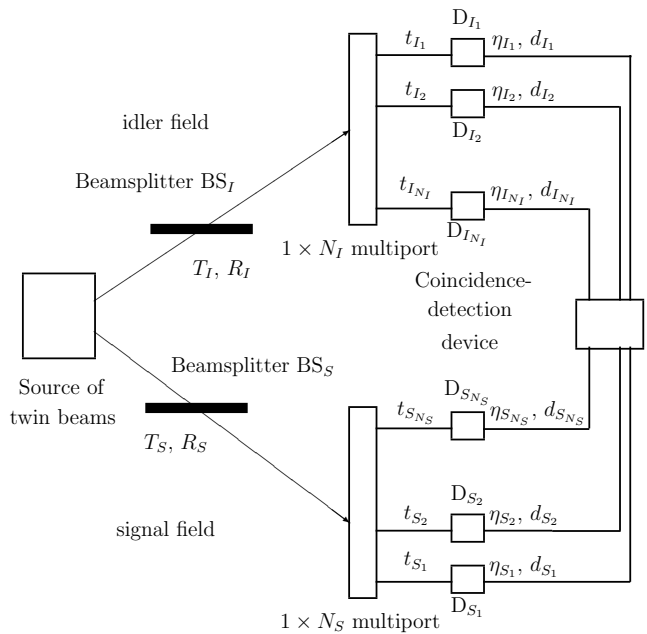}}
 \vspace{3mm}
  \caption{Scheme of the considered model.
  Photon pairs are generated in a nonlinear crystal
  NLC. Virtual beam-splitters $ {\rm BS}_S $ and $ {\rm BS}_I $
  describe possible losses of one or both photons from a pair before they are
  detected. Signal (idler) photons propagate through
  a $ 1\times N_S $ ($ 1\times N_I $) multi-port and are detected
  by single-photon detectors $ {\rm D}_{S_1} $, $ {\rm D}_{S_2} $, \ldots,
  $ {\rm D}_{S_{N_S}} $ ($ {\rm D}_{I_1} $, $ {\rm D}_{I_2} $, \ldots,
  $ {\rm D}_{I_{N_I}} $). Signals from the detectors are registered
  in a coincidence-detection device.}
\label{fig1}
\end{figure}

The measurement of joint signal-idler photon-number distribution
can be in general described using the scheme shown in
Fig.~\ref{fig1}
\cite{PerinaJr2003,PerinaJr2001,Haderka2001,Haderka2001a}.
Photon pairs occurring in the output plane of a nonlinear crystal
NLC propagate towards photon-number-resolving detection devices
placed in the paths of the signal and idler fields. One or both
photons from a pair may be lost before they reach their detection
devices. Geometric filtering (one photon from a pair is not
steered to the detector area), reflections on optical elements in
the experimental setup or absorption of a photon along its path to
a detection device represent possible reasons. We describe this
effect by two beam-splitters BS$ {}_S $ and BS$ {}_I $
\cite{Campos1989} placed in the signal- and idler-field paths,
respectively. We model a photon-number-resolving detection device
as a multi-port $ 1\times N $ \cite{Torma1996} followed by $ N $
single-photon detectors. This description holds also in the
special case when an intensified CCD camera is used
\cite{Jost1998}. We note that detectors able to resolve directly
photon numbers to some extent have been constructed
\cite{Kim1999,Miller2003,Brattke2001}. From practical viewpoint,
also detectors using time multiplexing (reached, e.g., in fiber
optics) and one or two single-photon detectors are promising
\cite{Rehacek2003,Achilles2003,Fitch2004}. Photon-number-resolving
detection in all these devises can be described in the framework
of the presented general model.

We assume that the signal and idler fields in the output plane of
the nonlinear crystal NLC are described by the following
statistical operator $ \hat{\rho}_{SI} $ written in the Fock
basis:
\begin{equation}           
 \hat{\rho}_{SI} = \sum_{n_S=0}^{\infty} \sum_{n_I=0}^{\infty}
  p(n_S,n_I) |n_S\rangle_S {}_S \langle n_S| \otimes
  |n_I\rangle_I {}_I \langle n_I| ;
\end{equation}
the symbol $ p(n_S,n_I) $ stands for the joint signal-idler
photon-number distribution.

Statistical operator $ \hat{\rho}^D_{SI} $ appropriate for the
signal and idler fields in front of the detection devices can be
expressed as \cite{PerinaJr2003}:
\begin{eqnarray}           
 \hat{\rho}^D_{SI} &=& \sum_{n_S=0}^{\infty} \sum_{n_I=0}^{\infty}
  p(n_S,n_I) \nonumber \\
  & & \mbox{} \times \sum_{l_S=0}^{n_S}
  \pmatrix{n_S \cr l_S\cr}  T_S^{l_S} R_S^{n_S-l_S}
|l_S\rangle_S {}_S \langle l_S| \nonumber \\
 & & \mbox{} \times \sum_{l_I=0}^{n_I}
  \pmatrix{n_I \cr l_I\cr}  T_I^{l_I} R_I^{n_I-l_I}
  |l_I\rangle_I {}_I \langle l_I|  .
  \label{2}
\end{eqnarray}
The symbols $ R_S $ and $ R_I $ ($ T_S $ and $ T_I $) denote
intensity reflectivities (transmissivities) of the beam-splitters
in the corresponding path.

We assume a multi-port $ 1\times N_S $ ($ 1\times N_I $) followed
by $ N_S $ ($ N_I $) single-photon detectors with quantum
efficiencies $ \eta_{S_j} $ ($ \eta_{I_j} $) and dark-count rates
$ d_{S_j} $ ($ d_{I_j} $) in the signal (idler) path. Detection of
a photon in the $ k $-th detector is described by the following
detection operator $ \hat{D}_{i_k} $ \cite{PerinaJr2001}:
\begin{eqnarray}       
 \hat{D}_{i_k} &=& \sum_{n=0}^{\infty} \{ [1- (1-\eta_{i_k})^n]
 + d_{i_k} (1-\eta_{i_k})^n \} |n\rangle_k {}_k \langle n| ,
 \nonumber \\
 & & i=S,I.
 \label{3}
\end{eqnarray}
On the other hand, detection operator $ \hat{D}_{i_k}^{\rm no} $
corresponds to the case when no detection has occurred:
\begin{equation}      
 \hat{D}_{i_k}^{\rm no} = 1 - \hat{D}_{i_k} .
\end{equation}
The effect of 'splitting' photons in the signal field in a $
1\times N_S $ multi-port can be described by the relation $
\hat{a}_S = \sum_{j=1}^{N_S} t_{S_j} \hat{a}_{S_j} $, where $
\hat{a}_S $ is the annihilation operator in the signal field
entering the multi-port, whereas the annihilation operator $
\hat{a}_{S_j} $ describes a field at the $ i $-th multi-port
output. Symbol $ t_{S_j} $ stands for an amplitude transmissivity
of a photon from the input to the $ j $-th output. The $ 1\times
N_I $ multi-port in the idler-field path is described similarly
and the symbol $ t_{I_j} $ then refers to an amplitude
transmissivity of a photon from the input to the $ j $-th
multi-port output.

The probability $ C_{{S^D},{I^D}} $ that given $ c_S $ detectors
in the signal field and given $ c_I $ detectors in the idler field
detect a photon whereas the rest of detectors does not register a
photon is determined as the quantum mean value:
\begin{eqnarray}     
C_{{S^D},{I^D}} &=& {\rm Tr}_{SI} \left\{ \hat{\rho}^D_{SI}
\prod_{a\in {S^D}} \hat{D}_a \; \prod_{b\in S \backslash {S^D}}
\hat{D}_b^{\rm no} \right. \nonumber \\
& &  \left. \times
\prod_{c\in {I^D}} \hat{D}_c \; \prod_{d\in I \backslash {I^D}}
\hat{D}_d^{\rm no} \right\}.
 \label{5}
\end{eqnarray}
The symbol $ S $ ($ I $) denotes the set of all signal-field
(idler-field) detectors $ S = \{S_1,\ldots,S_{N_S} \} $ ($ I =
\{I_1,\ldots,I_{N_I} \} $). The set $ S^D $ ($ I^D $) contains
signal-field (idler-field) detectors that have registered a
photon. Symbol $ {\rm Tr} $ stands for an operator trace.

Using the statistical operator $ \hat{\rho}^D_{SI} $ given in
Eq.~(\ref{2}) the probability $ C_{{S^D},{I^D}} $ of a
multi-coincidence count defined in Eq.~(\ref{5}) is obtained in
the form:
\begin{eqnarray}        
 C_{{S^D},{I^D}} &=& \sum_{n_S=0}^{\infty} \sum_{n_I=0}^{\infty}
  p(n_S,n_I) K_{S,{S^D}}(n_S) K_{I,{I^D}}(n_I) ,
   \nonumber \\
  K_{S,{S^D}}(n_S) &=& (-1)^{c_S} \left[ \prod_{b\in S}(1-d_b)
  \right] \nonumber \\
   & & \hspace{-1.7cm} \times \left[ T_S\left( \sum_{c\in S}
 |t_{c}|^2(1-\eta_c) \right) + R_S \right]^{n_S}  \nonumber \\
   & & \hspace{-1.7cm} \mbox{} +
   \frac{(-1)^{c_S-1}}{1!} \sum_{a\in {S^D}} \left[  \prod_{b\in S
   \backslash \{a\}} (1-d_b)  \right] \nonumber \\
   & & \hspace{-1.7cm} \times
  \left[ T_S\left( |t_a|^2 \eta_a + \sum_{c\in S\backslash \{a\}}
  |t_{c}|^2(1-\eta_c) \right) +  R_S \right]^{n_S}  \nonumber \\
   & & \hspace{-1.7cm} \mbox{} + \ldots
   + \left[ \prod_{b\in S\backslash {S^D}}(1-d_b)
   \right]  \nonumber \\
   & & \hspace{-1.7cm} \times
  \left[ T_S\left( \sum_{c\in {S^D}} |t_c|^2 + \sum_{c\in
   S\backslash {S^D}} |t_{c}|^2(1-\eta_c) \right) +
   R_S \right]^{n_S} , \nonumber \\
  K_{I,{I^D}}(n_I) &=& (-1)^{c_I} \left[ \prod_{b\in I}(1-d_b)
  \right] \nonumber \\
  & & \hspace{-1.7cm} \times
  \left[ T_I\left( \sum_{c\in I} |t_{c}|^2(1-\eta_c) \right) +
   R_I \right]^{n_I}  \nonumber \\
   & & \hspace{-1.5cm} \mbox{} +
   \frac{(-1)^{c_I-1}}{1!} \sum_{a\in {I^D}} \left[  \prod_{b\in I
   \backslash \{a\}} (1-d_b)  \right]  \nonumber \\
   & & \hspace{-1.7cm} \times
  \left[ T_I\left( |t_a|^2 \eta_a + \sum_{c\in I\backslash \{a\}}
  |t_{c}|^2(1-\eta_c) \right) +  R_I \right]^{n_I} \nonumber \\
   & & \hspace{-1.7cm} \mbox{} + \ldots
   + \left[ \prod_{b\in I\backslash {I^D}}(1-d_b)
   \right] \nonumber \\
   & & \hspace{-1.7cm} \times
  \left[ T_I\left( \sum_{c\in {I^D}} |t_c|^2 + \sum_{c\in
   I\backslash {I^D}} |t_{c}|^2(1-\eta_c) \right) +
   R_I \right]^{n_I} . \nonumber \\
  & &
  \label{6}
\end{eqnarray}

We now consider two symmetric multi-ports ($ t_{S_1}= t_{S_2}=
\ldots t_{S_{N_S}} = t_S = 1/\sqrt{N_S} $, $ t_{I_1}=t_{I_2}=
\ldots = t_{I_{N_I}} = t_I = 1/\sqrt{N_I} $) and detectors endowed
with the same characteristics in the signal and idler fields ($
\eta_{S_1} =\eta_{S_2} = \ldots = \eta_{S_{N_S}} = \eta_S $, $
d_{S_1} = d_{S_2} = \ldots = d_{S_{N_S}}= d_S $, $ \eta_{I_1} =
\eta_{I_2} = \ldots = \eta_{I_{N_I}} = \eta_I $, $ d_{I_1} =
d_{I_2} = \ldots = d_{I_{N_I}}= d_I $). Then the probability $
f^{N_S,N_I}(c_S,c_I) $ of having $ c_S $ detections somewhere at $
N_S $ signal detectors and $ c_I $ detections somewhere at $ N_I $
idler detectors can be expressed as:
\begin{equation}      
 f^{N_S,N_I}(c_S,c_I) = \pmatrix{N_S \cr c_S \cr}
 \pmatrix{N_I \cr c_I \cr} C_{{S^D},{I^D}} .
\end{equation}
Using the expression for $ C_{{S^D},{I^D}} $ provided in
Eq.~(\ref{6}) we arrive at the relation:
\begin{eqnarray}      
 f^{N_S,N_I}(c_S,c_I) &=& \sum_{n_S=0}^{\infty} \sum_{n_I=0}^{\infty}
 p(n_S,n_I) \nonumber \\
 & & \mbox{} \hspace{-3mm} \times  K^{S,N_S}(c_S,n_S) K^{I,N_I}(c_I,n_I) ,
 \label{8}
\end{eqnarray}
where
\begin{eqnarray}     
  K^{i,N_i}(c_i,n_i) &=& \pmatrix{N_i \cr c_i\cr} (1-d_i)^{N_i}
  (1-\tau_i)^{n_i} (-1)^{c_i} \nonumber \\
   & & \hspace{-1.5cm} \times \sum_{l=0}^{c_i}
  \pmatrix{c_i \cr l \cr} \frac{(-1)^l}{(1-d_i)^l}
  \left( 1 + \frac{l}{N_i} \frac{\tau_i}{1-\tau_i}
   \right)^{n_i} , \nonumber \\
   & & \hspace{1cm} i=S,I ;
 \label{9}
\end{eqnarray}
$ \tau_i $ ($ \tau_i = T_i \eta_i $) determines the probability
that a photon is registered at some of the detectors.

If the number of photons detected by the camera is much lower than
the number of pixels detecting the overall field with a
non-negligible probability, the limits $ N_S \longrightarrow
\infty $ and $ N_I \longrightarrow \infty $ are appropriate. When
determining these limits, the overall noise levels $ D_S $ and $
D_I $ are kept constant ($ D_S = N_S d_S $, $ D_I = N_I d_I $).
The coefficients $ K $ defined in Eq.~(\ref{9}) then considerably
simplify:
\begin{eqnarray}     
  K^{i,\infty}(c_i,n_i) &=& \sum_{l=0}^{\min (c_i,n_i)}
  \pmatrix{n_i \cr l\cr} (\tau_i)^l (1-\tau_i)^{n_i-l}
  \nonumber \\
  & & \mbox{} \hspace{-2mm} \times
   \frac{ D_i^{c_i-l}}{(c_i-l)!}  \exp(-D_i) , \hspace{0.3cm} i=S,I .
  \label{10}
\end{eqnarray}
We note that the following relations have been used when deriving
Eq.~(\ref{10}):
\begin{eqnarray}
 \sum_{k=0}^{N} \pmatrix{ N \cr k} (-1)^k (\alpha+k)^{n-1} &=& 0;
 \nonumber \\
 & & \mbox{} \hspace{-2cm}
  N\ge n \ge 1; 0^0\equiv 1; N,n\in N^+;
 \nonumber \\
 \sum_{k=0}^{N} \pmatrix{ N \cr k} (-1)^k (\alpha+k)^{N} &=&
 (-1)^N N!; \nonumber \\
 & & \mbox{} \hspace{-2cm} N\ge 0;0^0\equiv 1 ;
 \nonumber
\end{eqnarray}
symbol $ N^+ $ denotes positive integer numbers.

\section{Photon-number detection under real experimental
conditions}

In our typical experiment (see Fig.~\ref{fig2}) we define three
regions-of-interest on the iCCD detection photocathode: one for
collecting signal photons, one for counting idler photons and the
third one that serves for monitoring of the dark noise in the
experiment. To achieve higher data collection rates we use
hardware binning of several pixels to a single macro-pixel. The
signal and idler regions typically contain about one thousand of
macro-pixels that give an information about photons detection.
This means that a finite number of (macro-)pixels may play an
important role depending on intensity of the impinging field and
the general form of the transfer matrix $ K^{i,N_i}(c_i,n_i) $ in
Eq.~(\ref{9}) should be preferably used. However, evaluation of a
transfer matrix $ K $ for larger numbers of photons,
photoelectrons (registered photons) and (macro-)pixels is
difficult because a high extended precision in the evaluation of
the sum occurring in Eq.~(\ref{9}) is required. For instance, if
fields having up to 1000 photons are measured, from 2 to 3 hundred
significant decimal digits are needed in the evaluation of the
sums in Eq.~(\ref{9}) under conditions considered below. This is
time demanding and that is why we present several alternative ways
how to handle the problem under specific conditions.

First, we rewrite the relation between the measured frequencies $
f^{N_S,N_I} $ and the photon-number distribution $ p $ in a
general form:
\begin{eqnarray}      
 f^{N_S,N_I}(c_S,c_I) &=& \sum_{n_S=0}^{\infty} \sum_{n_I=0}^{\infty}
 p(n_S,n_I) \nonumber \\
 & & \mbox{} \hspace{-4mm} \times  G^{S,N_S}(c_S,n_S) G^{I,N_I}(c_I,n_I) ,
 \label{11}
\end{eqnarray}
where the general transformation matrices $ G^{i,N_i}(c_i,n_i) $
for $ i=S,I $ have been introduced.

In a real experimental setup, there are non-negligible losses
(described by intensity transmissivities $ T_{S} $ and $ T_I $)
before a field arrives to the photocathode of the iCCD camera. As
a result an average number of photons in the input to the camera
is lower compared to the average number of photons in the output
plane of the crystal. This may make a numerical evaluation of the
matrix $ K^{i,N_i} $ in Eq.~(\ref{9}) faster due to lower
dimensions of this matrix. In this case the transfer matrix $
G^{i,N_i}(c_i,n_i) $ can be rewritten as a product of two
matrices:
\begin{equation}   
 G^{i,N_i}(c_i,n_i) = \sum_{m=0}^{n_i} K^{i,N_i}(c_i,m)
 K_0(m,n_i).
\label{12}
\end{equation}
The matrix $ K_0 $ introduced in Eq.~(\ref{12}) describes the
Bernoulli distribution with transmissivity $ T_i $:
\begin{equation}   
 K_0(m,n_i) = \pmatrix{n_i \cr m} T_{i}^m(1-T_{i})^{n_i-m} .
 \label{13}
\end{equation}
Evaluation of the matrix $ K^{i,N_i} $ using Eq.~(\ref{9}) is then
done assuming $ \tau_i = \eta_i $.

If numbers of photons in a detected field are too high preventing
from the application of Eq.~({\ref{9}) (technical reasons in the
evaluation) we can proceed as follows. The measured field first
undergoes losses described by the intensity transmissivity $ T_{i}
$ before impinging on the camera. In the next step each photon
present in the region-of-interest of the camera 'registers' itself
in one (macro-)pixel. The probability $ \gamma^{i,N_i}(m_2,m_1) $
that $ m_1 $ photons register in $ m_2 $ (macro-)pixels assuming
the overall number of (macro-)pixels to be $ N_i $ is given by
permutations with repetition:
\begin{equation}   
 \gamma^{i,N_i}(m_2,m_1) = \frac{ \pmatrix{N_i \cr m_2}
 \pmatrix{m_1-1 \cr m_2-1} }{ \pmatrix{N_i+m_1-1 \cr N_i -1}},
  \hspace{4mm} m_2 \le N_i.
\end{equation}
In this case, $ m_2 $ (macro-)pixels are exposed by the field and
the probability of $ c_i $ detections ($ c_i \le m_2 $) is given
by the matrix $ K^{i,\infty}(c_i,m_2) $ written in Eq.~(\ref{10}),
i.e. as if there is an infinite number of (macro-)pixels in the
camera. The matrix $ G^{i,N_i}(c_i,n_i) $ then takes its final
approximative form:
\begin{eqnarray}   
 G^{i,N_i}(c_i,n_i) &=& \sum_{m_2=0}^{\min (m_1,N_i)} \sum_{m_1=0}^{n_i}
 K^{i,\infty}(c_i,m_2) \nonumber \\
 & & \mbox{} \times \gamma^{i,N_i}(m_2,m_1) K_0(m_1,n_i);
 \label{15}
\end{eqnarray}
the matrix $ K_0 $ is defined in Eq.~(\ref{13}).

It has been assumed in the derivation of Eq.~(\ref{15}) that each
of the exposed $ m_2 $ (macro)pixels contains only one photon (see
the limit $ N \longrightarrow \infty $). This approximation can be
improved. If $ m_1 $ photons is registered at $ m_2 $
(macro-)pixels, an average photon number occurring in one
(macro-)pixel is $ m_1/m_2 $. The average photon number $ m_1/m_2
$ greater than one leads to a higher probability of detection.
This increase of detection probability can be modelled by an
effective increase of detection quantum efficiency (see Appendix
A). The improved matrix $ G^{i,N_i}(c_i,n_i) $ can then be
determined along the relation:
\begin{equation}   
 G^{i,N_i}(c_i,n_i) = \sum_{m=0}^{n_i} \Gamma^{i,N_i}(c_i,m) K_0(m,n_i)
\end{equation}
and
\begin{eqnarray}   
 \Gamma^{i,N_i}(c_i,m) &=& \sum_{m_2=1}^{\min (m,N_i)} \sum_{l=0}^{\min (c_i,m_2)}
 \pmatrix{m_2 \cr l} \nonumber \\
 & & \mbox{} \hspace{-1.5cm}\times
 \left[1-\exp\left(-\eta_{i}\frac{m}{m_2}\right)\right]^l
 \left[\exp\left(-\eta_{i}\frac{m}{m_2}\right)\right]^{m_2-l}
 \nonumber \\
 & & \mbox{} \hspace{-1.5cm} \times
 \frac{D_i^{c_i-l}}{(c_i-l)!} \exp(-D_i)
 \frac{ \pmatrix{N_i \cr m_2}
 \pmatrix{m-1 \cr m_2-1} }{ \pmatrix{N_i+m-1 \cr N_i -1}}.
\end{eqnarray}

On the other hand weak detected fields allow the following
simplification. If the maximum number $ c_i $ of counts is much
less than the number of (macro-)pixels $ N_i $ the expression for
matrix $ K^{i,N_i} $ in Eq.~(\ref{9}) can be successfully
approximated using the relation $ (1+x)^{n_i} \approx \exp(xn_i) $
for $ |x| \ll 1 $. We then arrive at the matrix $
G^{i,N_i}(c_i,n_i) $ in the form:
\begin{equation}   
 G^{i,N_i}(c_i,n_i) = \sum_{m=0}^{n_i} K_{\rm exp}^{i,N_i}(c_i,m) K_0(m,n_i)
\end{equation}
and
\begin{eqnarray}   
 K_{\rm exp}^{i,N_i}(c_i,m) &=& \pmatrix{N_i \cr c_i}
 (1-d_i)^{N_i-c_i} (1-\eta_{i})^{m-c_i} \nonumber \\
 & & \mbox{} \hspace{-6mm} \times
  \left[ d_i (1-\eta_{i}) + \eta_{i} \frac{m}{N_i}
  \right]^{c_i} .
\label{19}
\end{eqnarray}
The expression in Eq.~(\ref{19}) has a simple interpretation: $
m-c_i $ impinging photons are not registered with probability $
1-\eta_{i} $ per photon. There occur $ c_i $ counts given either
by impinging photons with probability $ m\eta_{i}/N_i $ per photon
or by dark counts with probability $ d_i(1-\eta_{i}) $. $ N_i -
c_i $ (macro-)pixels cannot feel dark counts with probability $
1-d_i $ per (macro-)pixel.

\section{Inhomogeneous transverse intensity profile of a detected
beam}

If the intensity transverse profile of a beam impinging on an iCCD
camera is inhomogeneous, we can divide (macro-)pixels of the
camera into $ M_i $ groups assuming the same level of illumination
of (macro-)pixels belonging to one group. A $ j $-th group of
(macro-)pixels is characterized by probability $ \tau_{i_j} $ that
a photon present in beam $ i $ ($ i=S,I $) impinges on one
(macro-)pixel from this group, number $ \nu_{i_j} $ of
(macro-)pixels, quantum detection efficiency $ \eta_{i_j} $
dark-count rate $ d_{i_j} $, and mean number $ \mu_{i_j} $ of
photons impinging on one (macro-)pixel. It holds that $
\sum_{j=1}^{M_i} \tau_{i_j}\nu_{i_j} = \eta_i $ and $
\sum_{j=1}^{M_i} \nu_{i_j} = N_i $. The probability $ \tau_{i_j} $
that a photon reaches one (macro-)pixel of the $ j $-th group is
linearly proportional to the mean number $ \mu_{i_j} $ of photons
coming to this (macro-)pixel and can be expressed as:
\begin{equation}     
 \tau_{i_j} = \frac{\mu_{i_j}}{\mu_i^{\rm aver}N_i}, \hspace{5mm} i=S,I.
 \label{20}
\end{equation}
The average mean photon number $ \mu_{i}^{\rm aver} $ is given as
$ \mu_{i}^{\rm aver} = \sum_{j=1}^{M_i} \mu_{i_j} \nu_{i_j} / N_i
$.

A transformation matrix $ \tilde{K}^{i,N_i}(c_i,n_i) $ that
generalizes the matrix $ K^{i,N_i} $ occurring in Eq.~(\ref{9})
and determines the probability of $ c_i $ counts caused by $ n_i $
photons coming to the camera is given by the following $ M_i
$-dimensional convolution of matrices $ K^{i,\nu_{ij}} $ written
in Eq.~(\ref{9}) and characterizing the detection in the $ j $-th
group of (macro-)pixels:
\begin{eqnarray}  
 \tilde{K}^{i,N_i}(c_i,n_i) &=& \left\{ \left. \left[ \prod_{j=1}^{M_i}
 \sum_{n_j=0}^{n_i} \right] \right|_{\sum_{j=1}^{M_i}n_j=n_i} \right\}
 \nonumber \\
 & & \mbox{} \times
 \left\{ \left. \left[ \prod_{j=1}^{M_i}
 \sum_{c_j=0}^{c_i} \right] \right|_{\sum_{j=1}^{M_i}c_j=c_i} \right\}
 \nonumber \\
 & & \mbox{} \times
 n_i! \prod_{j=1}^{M_i} \frac{(\tau_{i_j}\nu_{i_j})^{n_j}}{n_j!}
 K^{i,\nu_{i_j}}(c_j,n_j) ,
 \nonumber \\
 & & \mbox{}  \hspace{3cm} i=S,I.
 \label{21}
\end{eqnarray}

The matrix $ \tilde{K}^{i,N_i} $ occurring in Eq.~(\ref{21}) can
be rewritten into the following form if the matrices $
K^{i,\nu_{i_j}} $ are expressed using the relation in
Eq.~(\ref{9}):
\begin{eqnarray}  
 \tilde{K}^{i,N_i}(c_i,n_i) &=& \left\{ \left. \left[ \prod_{j=1}^{M_i}
 \sum_{c_j=0}^{c_i} \right] \right|_{\sum_{j=1}^{M_i}c_j=c_i} \right\}
 \nonumber \\
 & & \mbox{} \hspace{-22mm} \times
  \left[ \prod_{k=1}^{M_i} \pmatrix{ \nu_{i_k} \cr c_{k}}
  (1-d_{i_k})^{\nu_{i_k}} \right] \nonumber \\
 & & \mbox{} \hspace{-22mm} \times
  \left\{ \prod_{j=1}^{M_i} \sum_{l_j=0}^{c_j} \right\}
  \left[ \prod_{k=1}^{M_i} \pmatrix{c_k \cr l_k}
  \frac{(-1)^{l_k}}{(1-d_{i_k})^{l_k}} \right]
  \nonumber \\
 & & \mbox{} \hspace{-22mm} \times (-1)^{c_i}
  \left[ 1 - \sum_{k=1}^{M_i} (\tau_{i_k}\nu_{i_k}\eta_{i_k})
   + \sum_{k=1}^{M_i} (l_{k}\tau_{i_k}\eta_{i_k}) \right]^{n_i} .
  \label{22}
\end{eqnarray}

If the number of (macro-)pixels is sufficiently large compared to
the number of impinging photons, consideration of the following
limit is useful. In this limit $ \nu_{i_j} \rightarrow \infty $
for $ j=1,\ldots,M_i $ assuming $ \nu_{i_j}\tau_{i_j} $
[probability that a photon is detected in the $ j $-th group of
(macro-)pixels] to be constant. Also $ d_{i_j}\nu_{i_j} = D_{i_j}
$ [overall dark-count rate of all (macro-)pixels in the $ j $-th
group] is assumed to be constant. Then the expression in
Eq.~(\ref{22}) simplifies and leaves the matrix $
\tilde{K}^{i,N_{i}} $ in the form:
\begin{eqnarray}  
 \tilde{K}^{i,\infty}(c_i,n_i) &=& \left\{ \left. \left[ \prod_{j=1}^{M_i}
 \sum_{c_j=0}^{c_i} \right] \right|_{\sum_{j=1}^{M_i}c_j=c_i} \right\}
 \nonumber \\
 & & \mbox{} \hspace{-20mm} \times
  \left\{\prod_{j=1}^{M_i} \sum_{l_j=0}^{\min(c_j,n_i)} \right\}
  \frac{n_i!}{\left[ \prod_{k=1}^{M_i} l_{k}! \right]
  \left( n_i-\sum_{k=1}^{M_i} l_{k} \right)! }
  \nonumber \\
  & & \mbox{} \hspace{-20mm} \times
  \left[ \prod_{k=1}^{M_i}(\tau_{i_k}\nu_{i_k}\eta_{i_k})^{l_k}\right]
  \left[ \prod_{k=1}^{M_i} \frac{ D_{i_k}^{c_k-l_k} }{(c_k-l_k)!}
  \exp(-D_{i_k}) \right] \nonumber \\
  & & \mbox{} \hspace{-20mm} \times
  \left[ 1- \sum_{k=1}^{M_i} (\tau_{i_k}\nu_{i_k}\eta_{i_k})
  \right]^{n_i-\sum_{k=1}^{M_i} l_k} , \hspace{3mm} i=S,I.
 \label{23}
\end{eqnarray}
The expression in Eq.~(\ref{23}) has a simple interpretation: $
n_i $ photons impinging on the camera generates $ l_j $ counts in
a $ j $-th group of (macro-)pixels with probability $
\tau_{i_j}\nu_{i_j}\eta_{i_j} $ per photon and $ (c_j - l_j) $
counts come from dark counts occurring in the $ j $-th group of
(macro-)pixels. The remaining $ n_i -\sum_{k=1}^{M_i} l_k  $
photons are not registered with probability $ 1- \sum_{j=1}^{M_i}
(\tau_{i_j}\nu_{i_j}\eta_{i_j}) $ per photon.

Provided that the maximum number of counts $ c_j $ in a $ j $-th
group of (macro-)pixels is much less than the number of
(macro-)pixels $ \nu_{i_j} $ in this group the approximate
relation $ (1+x)^{n_i} \approx \exp(n_i x) $ for $ |x| \ll 1 $
enables to rearrange the formula in Eq.~(\ref{22}) as follows:
\begin{eqnarray}  
 \tilde{K}^{i,N_i}_{\rm exp}(c_i,n_i) &=& \left\{ \left. \left[ \prod_{j=1}^{M_i}
 \sum_{c_j=0}^{c_i} \right] \right|_{\sum_{j=1}^{M_i}c_j=c_i} \right\}
 \nonumber \\
  & & \mbox{} \hspace{-20mm} \times
  \left(
  1-\sum_{k=1}^{M_i}(\tau_{i_k}\nu_{i_k}\eta_{i_k})\right)^{n_i - c_i}
   \nonumber \\
 & & \mbox{} \hspace{-20mm} \times
  \Biggl[ \prod_{j=1}^{M_i} \pmatrix{ \nu_{i_j} \cr c_j }
  (1-d_{i_j})^{\nu_{i_j}-c_j} \nonumber \\
  & & \hspace{-20mm} \times
  \left(
   d_{i_j}\left[1-\sum_{k=1}^{M_i}(\tau_{i_k}\nu_{i_k}\eta_{i_k})\right]
   + n_i\tau_{i_j}\eta_{i_j} \right)^{c_j} \Biggr] .
  \label{24}
\end{eqnarray}
The expression in Eq.~(\ref{24}) can be interpreted similarly as
the formula in Eq.~(\ref{19}). If $ c_i $ counts occur after $ n_i
$ photons enter the camera, $ n_i - c_i $ photons is not
registered with probability $ 1-\sum_{k=1}^{M_i}
\tau_{i_k}\nu_{i_k}\eta_{i_k} $ per photon. In a $ j $-th group of
(macro)pixels $ \nu_{i_j}-c_j $ (macro-)pixels do not count a
photon with probability $ 1- d_{i_j} $ per (macro-)pixel (dark
counts have to be `eliminated'). Finally $ c_j $ (macro-)pixels
detect a photon either due to a successful registration of a
photon taken from $ n_i $ incident photons with probability $
\tau_{i_j}\eta_{i_j} $ per photon or owing to a dark count with
probability $ d_{i_j}(1-\sum_{k=1}^{M_i}
\tau_{i_k}\nu_{i_k}\eta_{i_k} )$ (a dark count occurs if there is
no detection caused by an impinging photon).

If the number of counts $ c_i $ registered by an iCCD camera is
low and the number of groups of (macro-)pixels is greater, a
useful alternative expression for the transfer matrix $
\tilde{K}^{i,N_i}(c_i,n_i) $ given in Eq.~(\ref{21}) can be
derived directly from Eq.~(\ref{6}) ($ T_i=1 $ and $ R_i=0 $ are
assumed, $ i=S,I $):
\begin{eqnarray}  
 \tilde{K}^{i,N_i}(c_i,n_i) &=& \left\{ \left. \left[ \prod_{j=1}^{M_i}
 \sum_{c_j=0}^{c_i} \right] \right|_{\sum_{j=1}^{M_i}c_j=c_i} \right\}
 \nonumber \\
 & & \mbox{} \hspace{-10mm}\times
 \tilde{K}_{i,\{c_{1},\ldots,c_{M_i}\}} (c_i,n_i) , \hspace{5mm} i=S,I,
\end{eqnarray}
where the coefficient $
\tilde{K}_{i,\{c_{1},\ldots,c_{M_i}\}}(c_i,n_i) $ gives the
probability that $ c_j $ counts have occurred in the $ j $-th
group of (macro-)pixels; $ \sum_{j=1}^{M_i} c_j = c_i $. It can be
expressed as follows:
\begin{eqnarray}  
 \tilde{K}_{i,\{c_{1},\ldots,c_{M_i}\}}(c_i,n_i) &=& \left[ \prod_{j=1}^{M_i}
 \pmatrix{\nu_{i_j} \cr c_j}(1-d_{i_j})^{\nu_{i_j}} \right]
 \nonumber \\
 & & \mbox{} \hspace{-35mm} \Biggl\{
 (-1)^{c_i} \Theta_i^{n_i} +
 (-1)^{c_i-1} \sum_{k_1=1}^{c_i} \frac{\left[
 \Theta_i + \tau_{i\sigma_i(k_1)}\eta_{i\sigma_i(k_1)} \right]^{n_i}
 }{1-d_{i\sigma_i(k_1)}}
 \nonumber \\
 & & \mbox{} \hspace{-35mm} +
 (-1)^{c_i-2} \sum_{k_1=1}^{c_i-1} \sum_{k_2=k_1+1}^{c_i}
 \nonumber \\
 & & \mbox{} \hspace{-25mm} \frac{ \left[
  \Theta_i + \tau_{i\sigma_i(k_1)}\eta_{i\sigma_i(k_1)} +
  \tau_{i\sigma_i(k_2)}\eta_{i\sigma_i(k_2)} \right]^{n_i}
   }{(1-d_{i\sigma_i(k_1)})(1-d_{i\sigma_i(k_2)})}
 \nonumber \\
 & & \mbox{} \hspace{-35mm} +
 (-1)^{c_i-3} \sum_{k_1=1}^{c_i-2} \sum_{k_2=k_1+1}^{c_i-1} \sum_{k_3=k_2+1}^{c_i}
 \nonumber \\
 & & \mbox{} \hspace{-25mm}
  \frac{\left[
  \Theta_i + \sum_{m=1}^{3}
  (\tau_{i\sigma_i(k_m)}\eta_{i\sigma_i(k_m)}) \right]^{n_i}
  }{ \prod_{m=1}^{3} (1-d_{i\sigma_i(k_m)}) } +
  \ldots
 \nonumber \\
 & & \mbox{} \hspace{-35mm} +
 \frac{ \left[
  \Theta_i + \sum_{m=1}^{c_i}
  (\tau_{i\sigma_i(k_m)}\eta_{i\sigma_i(k_m)}) \right]^{n_i}
  }{ \prod_{m=1}^{c_i} (1-d_{i\sigma_i(k_m)}) }
  \Biggr\} .
\label{26}
\end{eqnarray}
The symbol $ \Theta_i $ introduced in Eq.~(\ref{26}) denotes the
probability that a photon is not registered by the camera; i.e.
\begin{equation}   
 \Theta_i = 1- \sum_{j=1}^{M_i} (\tau_{i_j}\nu_{i_j}\eta_{i_j}) , \hspace{5mm} i=S,I.
\end{equation}
The vector $ \sigma_i $ in Eq.~(\ref{26}) is composed of $ c_i $
elements ($ i=S,I $); its $ j $-th element gives the number of
group of (macro-)pixels that registered the $ j $-th click ($
j=1,\ldots,c_i $). Thus, the first $ c_1 $ elements equal 1, the
next $ c_2 $ elements equal 2 and so on.

\section{Reconstruction of the joint signal-idler photon-number
distribution}

The probabilities (frequencies) $ f(c_S,c_I) $ are measured in the
experiment and the relation in Eq.~(\ref{11}) has to be inverted
in order to obtain the joint signal-idler photon-number
distribution $ p(n_S,n_I) $ beyond the nonlinear crystal. The
relation in Eq.~(\ref{11}) together with the coefficients $
G^{i,N_i}(c_i,n_i) $ defined in Eq. (\ref{12}) can be inverted
under special conditions analytically. For instance, if $ D_S =
D_I = 0 $ the inversion relation is found using the `convolution'
of function $ f $ with the Bernoulli distributions with
efficiencies $ 1/(T_S\eta_S) $ and $ 1/(T_I\eta_I) $ that are
greater than one. For more general cases, a method of direct
matrix inversion has been elaborated \cite{Waks2004,Avenhaus2008}.
However, analytical approaches are not suitable for processing
real experimental data \cite{Mogilevtsev1997} because of numerical
instabilities and the occurrence of artifacts. On the other hand
reconstruction algorithms have occurred to be suitable for this
task \cite{Rehacek2003}. Such algorithms are able to find a
reconstructed joint signal-idler photon-number distribution $
\rho_{\rm rec}(n_S,n_I) $ that matches the measured frequencies $
f(c_S,c_I) $ in the best way with respect to a given criterion.
Here, we consider the Kullback-Leibler divergence as a measure of
the distance between the experimental data and data provided by
the developed theory. The reconstructed joint signal-idler
photon-number distribution $ \rho_{\rm rec}(n_S,n_I) $ minimizing
the Kullback-Leibler divergence can then be found as a steady
point of an iteration algorithm \cite{Dempster1977,Vardi1993}:
\begin{eqnarray}      
\rho^{(n+1)}(n_S,n_I) &=& \rho^{(n)}(n_S,n_I)
  \nonumber \\
 & & \hspace{-2.8cm} \times \sum_{i_S,i_I=0}^{\infty}
  \frac{ f(i_S,i_I) G^{S,N_S}(i_S,n_S) G^{I,N_I}(i_I,n_I) }{
  \sum_{j_S,j_I=0}^{\infty} G^{S,N_S}(i_S,j_S)
  G^{I,N_I}(i_I,j_I) \rho^{(n)}(j_S,j_I) } .
 \nonumber \\
 & &
\label{28}
\end{eqnarray}
The symbol $ \rho^{(n)}(n_S,n_I) $ denotes a joint signal-idler
photon-number distribution after an $ n $-th step of the
iteration, $ \rho^{(0)}(n_S,n_I) $ is an arbitrary initial
photon-number distribution. We note that each element of the
initial photon-number distribution has to be nonzero in order to
be considered in the iteration process.

Convergence of the iteration process can be monitored using
parameter $ S $ that gives the declination of the reconstructed
photon-number distribution from the measured frequencies $
f(c_S,c_I) $ and is determined along the formula:
\begin{eqnarray}      
 S^{(n)} &=& \Biggl[ \sum_{c_S,c_I=0}^{\infty} \Bigl|
  f(c_S,c_I) \nonumber \\
  & & \hspace{-1cm} - \sum_{j_S,j_I=0}^{\infty} G^{S,N_S}(c_S,j_S)
   G^{I,N_I}(c_I,j_I) \rho^{(n)}(j_S,j_I) \Bigr|^2 \Biggr]^{1/2} .
   \nonumber \\
 & &
\label{29}
\end{eqnarray}

Alternatively, also covariance $ C $ of the signal and idler
photon numbers $ n_S $ and $ n_I $ derived for the joint
photon-number distribution $ \rho $,
\begin{eqnarray}          
  C^{(n)} &=& \frac{ \langle \Delta n_S \Delta n_I \rangle }{
  \sqrt{ \langle (\Delta n_S)^2
    \rangle \langle (\Delta n_I)^2 \rangle }} ,
      \label{30} \\
    & & \Delta n_i = n_i - \langle n_i \rangle,
      \hspace{0.5cm} i= S, I, \nonumber\\
   \langle n_S^k n_I^l \rangle &=& \sum_{n_S=0}^{\infty} \sum_{n_I=0}^{\infty}
     n_S^k n_I^l \rho^{(n)}(n_S,n_I) , \nonumber \\
    & &  \hspace{1cm} k,l=0,1,\ldots, \nonumber
\end{eqnarray}
can be used as a useful indicator. The reason is that the initial
photon-number distribution $ \rho^{(0)} $ is usually considered
without any correlation and the iteration process gradually
reveals photon-number correlations present in the joint signal and
idler field. We note that we have checked by numerical simulations
that the reconstruction algorithm cannot reveal correlations
provided that the measured frequencies $ f(c_S,c_I) $ describe two
independent fields.

Here, we analyze three photon-number distributions obtained under
different conditions using the experimental setup shown in
Fig.~\ref{fig2}.
\begin{figure}          
 \resizebox{0.9\hsize}{!}{\includegraphics{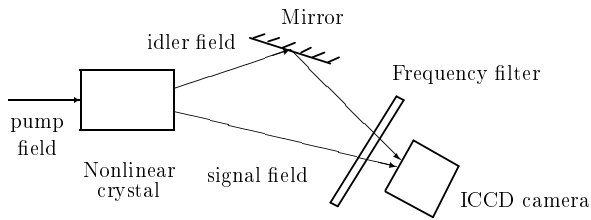}}
 \vspace{3mm}
  \caption{Scheme of the setup for detection of photon pairs.
  Fields composed of typically tens or hundreds of photon pairs are generated
  in a nonlinear crystal. Idler photons are reflected on a mirror. Both
  signal and idler photons propagate through a frequency
  filter and are detected in an iCCD camera.}
\label{fig2}
\end{figure}
Photon pairs are generated in a 5-mm long BBO crystal cut for a
type I process ($\theta=50$~deg, $\phi=90$~deg) pumped by
ultrashort pulses delivered by a cavity-dumped titanium-sapphire
femtosecond laser at the wavelength of 840~nm followed by a
third-harmonic generator. The pulses at the fundamental wavelength
are about 150~fs long. The laser system runs at the repetition
rate of 50~kHz and, after converting the 840-nm beam to its third
harmonic, it typically delivers pulses with the energy up to
45~nJ. Degenerate signal and idler photons occur at the cone layer
behind the crystal and leave the crystal at the outer output
half-angle of 13 degrees. Photons in the idler field are reflected
from a high-reflectivity mirror (>99~\% at 560~nm) and impinge on
an intensified CCD camera (Andor iStar 734). The camera has one
megapixel resolution with $13 \times 13$ $\mu$m$^2$ pixels but we
use $8 \times 8$ hardware binning to increase the readout rate.
The detection events are processed using our own software
employing thresholding, event centroid finding and photon
counting. The software is optimized to achieve maximum detection
efficiency which we evaluate to 23\% near the degenerate
wavelength of 560~nm. Three regions of interest are defined in the
field of view of the camera: the first one is for the signal
field, the second one for the idler field, and the last region
serves for reference measurements of the noise level. The whole
field of the camera is filtered by a high-transmittance high-pass
filter blocking light below 490~nm and an interference filter of
14~nm FWHM centered at 560~nm. The interference filter selects
nearly frequency degenerate photon pairs. Since a single run of
data acquisition usually takes several hours, the laser intensity
is actively stabilized (the intensity noise lays below 0.3~\% rms)
using feedback loop and polarization attenuator. The intensifier
of the camera is synchronously gated (gate duration equals 5~ns)
by cavity-dumper trigger pulses to minimize the noise from the
laboratory.

In the experimental setup, histograms $ f(c_S,c_I) $ of
photoelectron numbers have been taken under two different
intensity conditions. In the first case (a), the measurement has
been performed for lowest signal and idler intensities allowed by
the setup. The limiting intensities are given by noise of the
camera and stray light from the laboratory. The second case (b)
represents a typical result obtained under most suitable
conditions. The third case (c) corresponds to the measurement done
with greater signal and idler intensities and the histogram $
f(c_S,c_I) $ has been obtained by summing up five neighbor frames
together. We thus have three representative data sets with mean
photoelectron numbers equal to 1.2, 8.6, and 43. The corresponding
histograms $ f(c_S,c_I) $ are shown in Fig.~\ref{fig3}.
\begin{figure}      
 \vbox{\noindent (a)
  \resizebox{0.8\hsize}{!}{\includegraphics{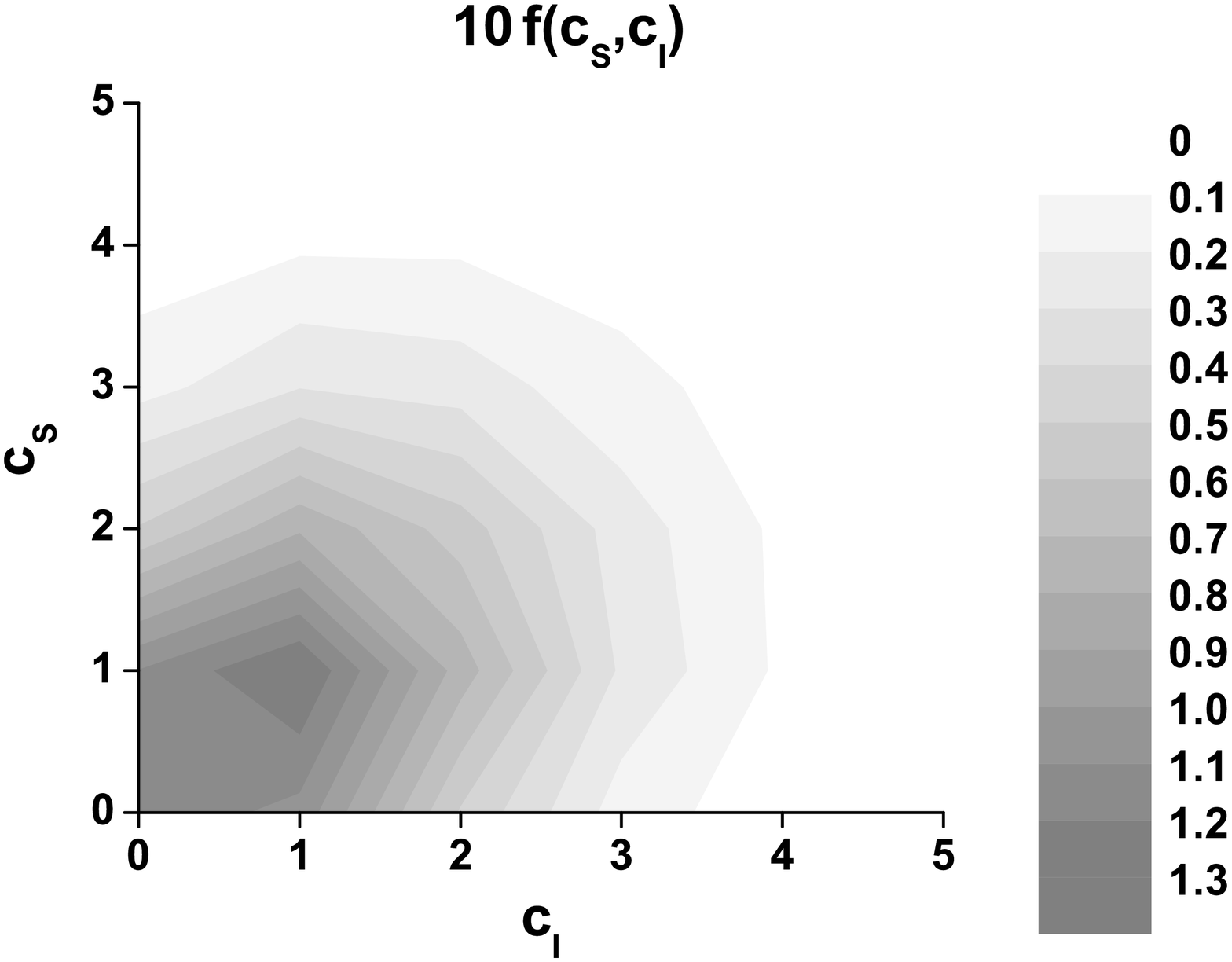}}
  \vspace{3mm}

 (b)
  \resizebox{0.8\hsize}{!}{\includegraphics{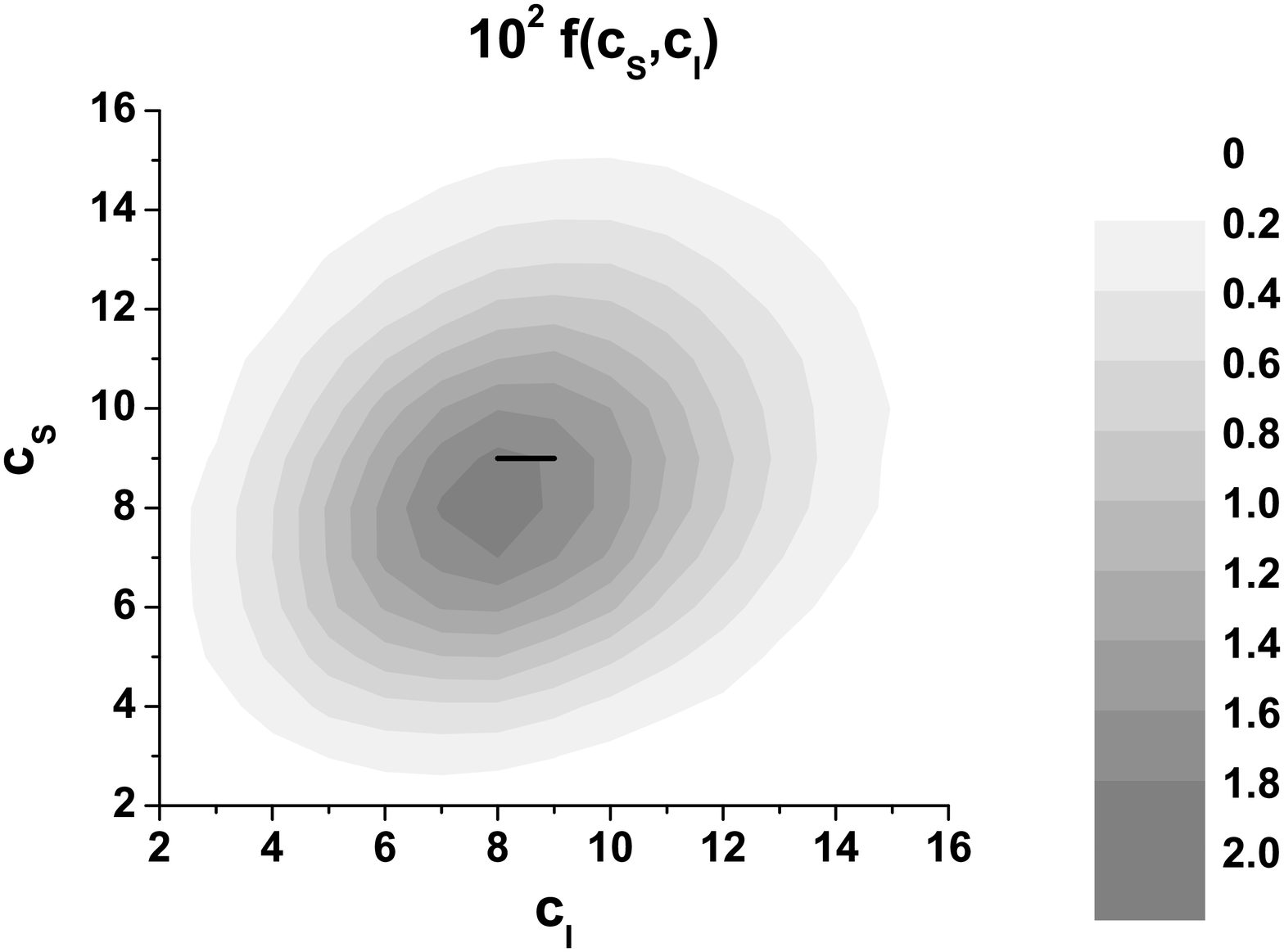}}
  \vspace{3mm}

 (c)
  \resizebox{0.8\hsize}{!}{\includegraphics{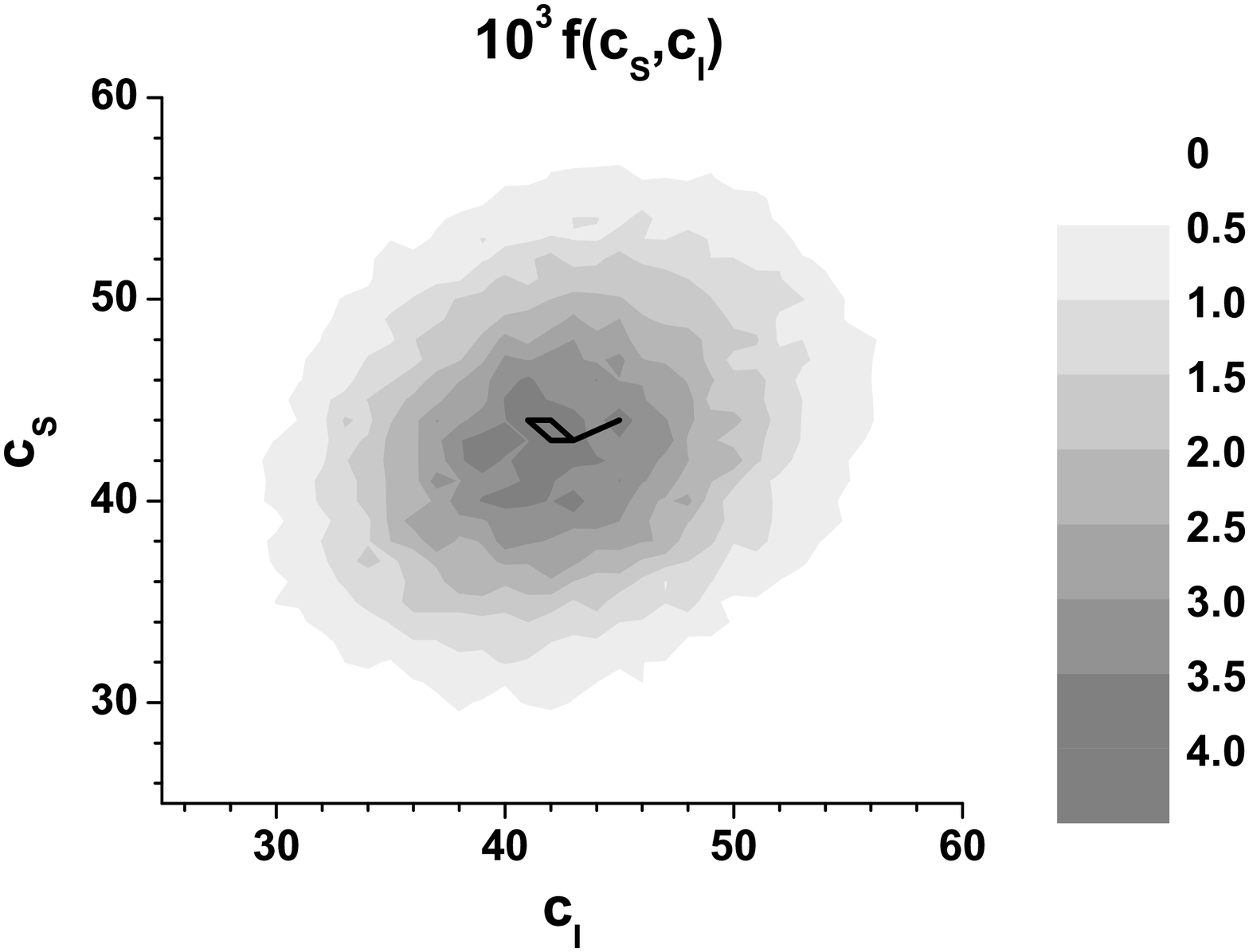}}  }
 \vspace{3mm}
  \caption{Topological graphs of the measured histograms $ f $ of
  signal ($ c_S $) and idler ($ c_I $) photoelectron numbers for three
  data sets denoted as (a), (b), and (c). A black curve
  encircles the area in which the classical inequality (\ref{32}) is violated.}
\label{fig3}
\end{figure}
Covariances of the photoelectron numbers $ c_S $ and $ c_I $
described by the histograms $ f(c_S,c_I) $ plotted in
Fig.~\ref{fig3} are in turn 23.8, 21.4, and 21.1~\%. This
corresponds to the expected overall detection efficiencies $
T_S\eta_S $ and $ T_I\eta_I $ around 20~\% and the low level of
single-photon noise.

In order to apply the iteration reconstruction algorithm described
in Eq.~(\ref{28}) we need to know the overall detection
efficiencies $ T_S\eta_S $ and $ T_I\eta_I $. In principle, they
can be determined by knowing parameters of the experimental setup.
However, fragility of the experimental alignment enforces the
determination from the obtained experimental data. The values of
detection efficiencies can either be derived from the measured
covariance between the signal and idler photoelectron numbers $
c_S $ and $ c_I $ or alternatively by applying a method described
in Sec.~VI below that relies on finding the best fit to the
experimental data. This method applied to the set of data (b) has
provided $ T_S\eta_S = 0.207 $ and $ T_I\eta_I = 0.205 $ that have
been used in the reconstruction. These values take into account
quantum detection efficiency of the iCCD camera as well as losses
occurring in the setup (frequency filters, reflection on the
output plane of the crystal and mirror). The reconstruction
algorithm also needs the level of dark noise that has been
monitored in the third region-of-interest of the photocathode; $
D_S = D_I = 0.03 $ (a), $ 0.09 $ (b), and $ 0.46 $ (c). The
application of the formula in Eq. (\ref{28}) has then resulted in
the joint signal-idler photon-number distributions $ p_{\rm
rec}(n_S,n_I) \equiv \rho^{(\infty)}(n_S,n_I) $ appropriate for
the output plane of the crystal and shown in Fig.~\ref{fig4}. The
initial joint signal-idler photon-number distribution $ \rho^{(0)}
$ has been taken as uniform in all three cases.
\begin{figure}      
 \vbox{\noindent (a)
 \resizebox{0.8\hsize}{!}{\includegraphics{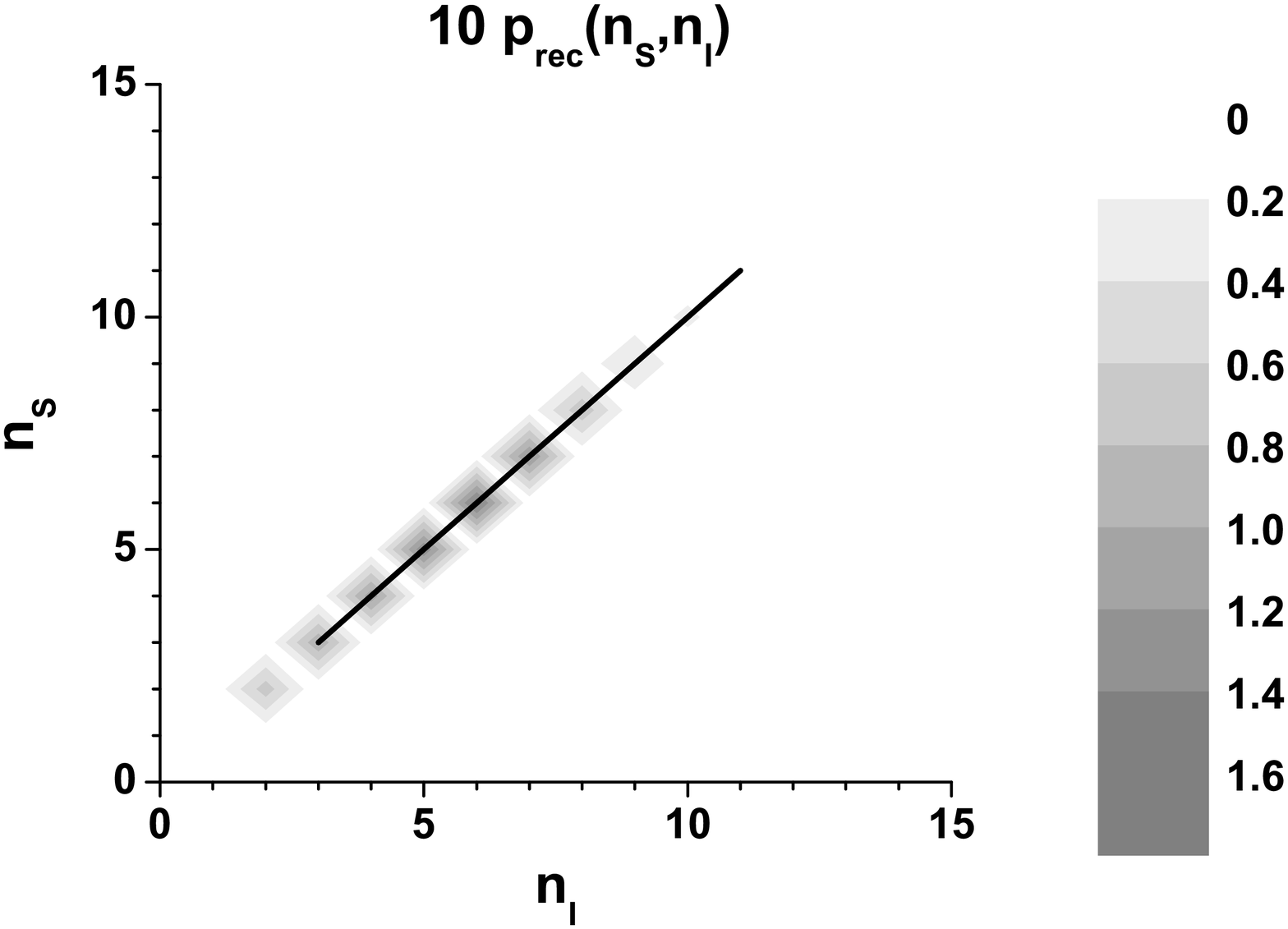}}
  \vspace{3mm}

 (b)
 \resizebox{0.8\hsize}{!}{\includegraphics{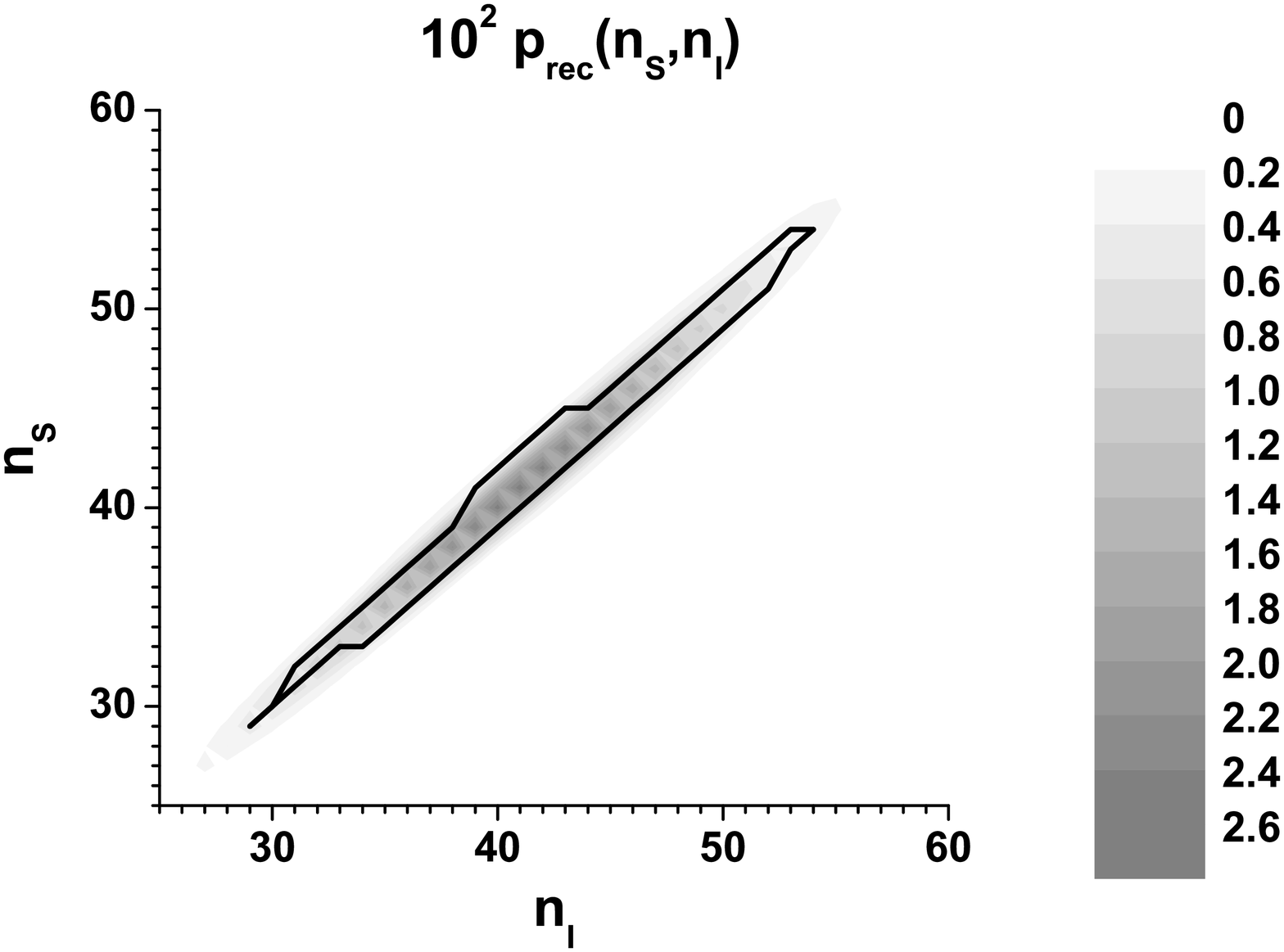}}
  \vspace{3mm}

 (c)
 \resizebox{0.8\hsize}{!}{\includegraphics{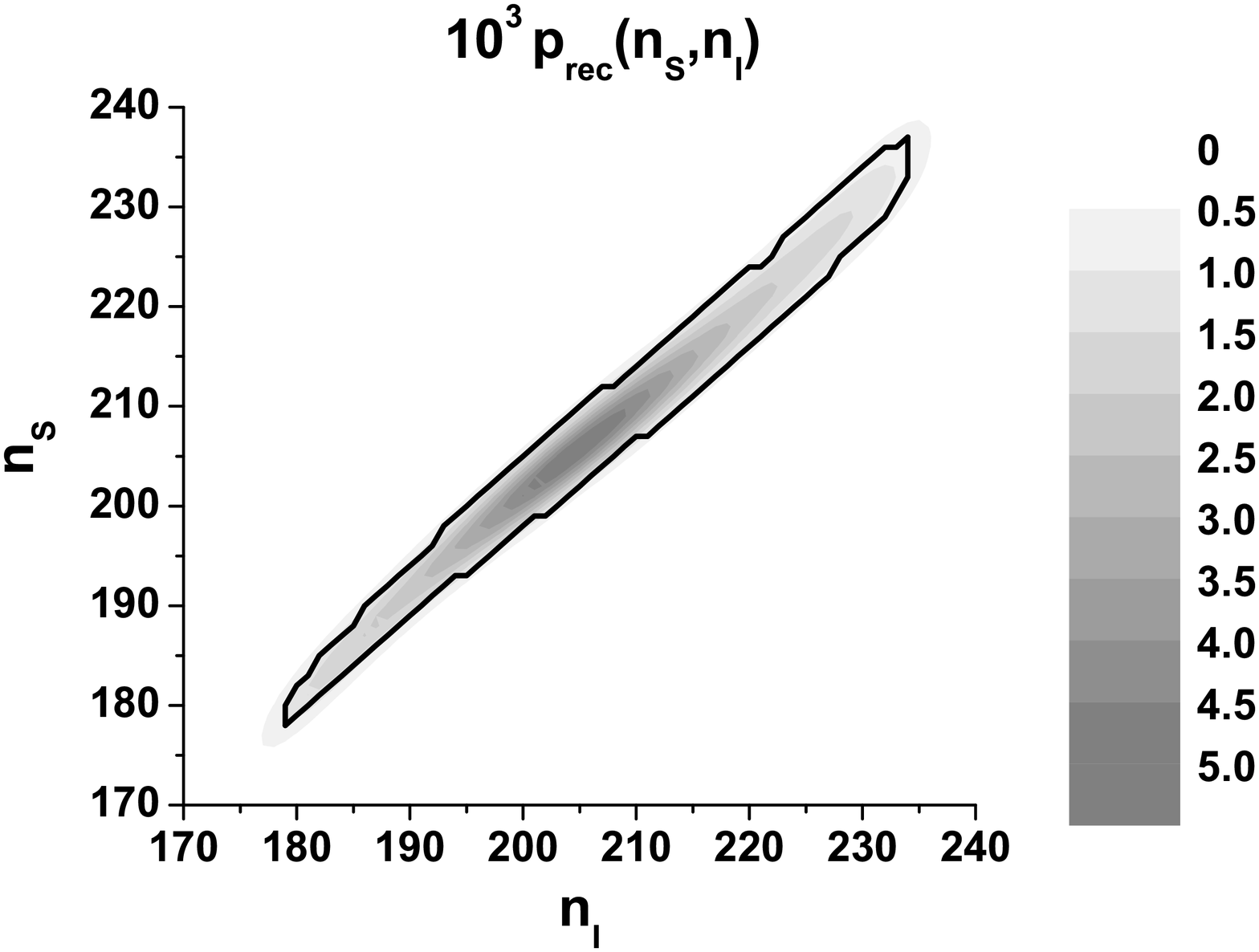}}
   }
 \vspace{3mm}
  \caption{Topological graphs of the reconstructed joint signal-idler photon-number
  distributions $ p_{\rm rec}(n_S,n_I) $ for data sets (a), (b), and (c). A black curve
  encircles the area in which the classical inequality (\ref{32}) is violated.}
\label{fig4}
\end{figure}
Covariances of the reconstructed joint signal-idler photon-number
distributions $ p_{\rm rec} $ equal 90.0~\% (a), 90.1~\% (b), and
89.7~\% (c). These numbers show the ability of the reconstruction
algorithm to recover paired correlations that have been weakened
during the propagation and detection process. The fact that the
obtained covariances do not approach 100~\% has two reasons: 1)
imperfect description of all noises occurring in the experiment,
2) numerical implementation of the iteration reconstruction
algorithm loses its precision with the increasing number of steps
\cite{Zambra2006}. Both values of the parameter $ S $ given in
Eq.~(\ref{29}) and covariance $ C $ defined in Eq.~(\ref{30}) can
be used for monitoring convergence of the iteration process.
Covariance $ C $ has been found to be more sensitive. Typically
several hundreds of iteration steps are needed to arrive at solid
(asymptotic) results as documented in Fig.~\ref{fig5} valid for
data set (b). In all three cases, 10 000 iteration steps have been
applied.
\begin{figure}      
 \resizebox{0.8\hsize}{!}{\includegraphics{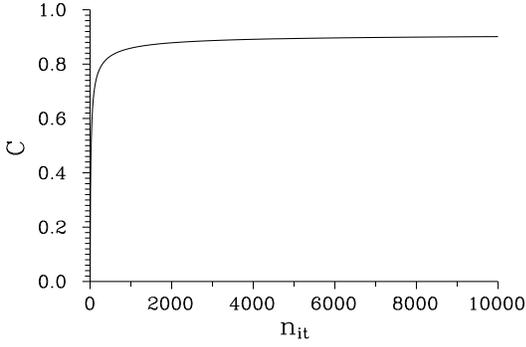}}
 \vspace{3mm}
  \caption{Covariance $ C $ of the signal and idler photon numbers
  $ n_S $ and $ n_I $ as it depends on the number $ n_{\rm it} $ of the
  iteration step for data set (b).}
\label{fig5}
\end{figure}

The reconstructed joint signal-idler photon-number distributions $
p_{\rm rec} $ show that the emitted fields are mainly composed of
photon pairs that are responsible for nonzero covariances of the
signal and idler photon numbers. Such fields are nonclassical in
the sense that they cannot be described by nonnegative
Glauber--Sudarshan quasi-distributions \cite{Perina1991}. As a
consequence, there even exist elements $ p_{\rm rec}(n_S,n_I) $ of
the joint photon-number distribution $ p_{\rm rec} $ that violate
the classical inequality \cite{Hillery1987}
\begin{equation}  
 p_{\rm rec}(n_S,n_I) \le \frac{\langle n_S\rangle^{n_S} \langle
  n_I\rangle^{n_I}}{n_S! \, n_I!} \exp[-\langle n_S\rangle
  -\langle n_I\rangle ].
\label{31}
\end{equation}

However, nonclassical properties manifest also in quantities
which determination is based on all elements of the joint
photon-number distribution $ p_{\rm rec} $.

\subsection{Important nonclassical characteristics of paired
fields}

Correlations in the signal and idler photon numbers $ n_S $ and $
n_I $ lead to narrowing of the distribution $ p_{-} $ of the
photon-number difference $ n_S - n_I $ together with broadening of
the distribution $ p_{+} $ of the photon-number sum $ n_S + n_I $.
The distributions $ p_{+} $ and $ p_{-} $ are defined as
\begin{eqnarray}         
 p_{\pm}(n) &=& \sum_{n_S=0}^{\infty} \sum_{n_I=0}^{\infty}
   \delta_{n,n_S \pm n_I} p_{\rm rec}(n_S,n_I) .
\label{32}
\end{eqnarray}
The symbol $ \delta $ stands for Kronecker's delta.

Fluctuations described by the distribution $ p_{-} $ of the
photon-number difference can even be lower than those
corresponding to any classical field with no correlations (see
Fig.~\ref{fig6}a). We then speak about sub-shot-noise correlations
that can be quantified by parameter $ R $ ($ R < 1 $ for
nonclassical states):
\begin{equation}   
 R = \frac{\langle \left[ \Delta(n_S-n_I)\right]^2 \rangle }{
  \langle n_S \rangle + \langle n_I \rangle }.
\label{33}
\end{equation}
Considering the experimental data, we obtain $ R= 0.133 $ (8.7~dB)
(a), $ R= 0.111 $ (9.5~dB) (b), and $ R= 0.117 $ (9.3~dB) (c).
This means that all three detected fields are strongly
nonclassical. We note that the discussed narrowing can also be
observed for stronger fields utilizing correlations of
photocurrents from two detectors \cite{Zhang2002}.

On the other hand, the distribution $ p_{+} $ of the photon-number
sum is super-Poissonian, i.e. its Fano factor F is greater than 1
[$ F = \langle (\Delta n)^2 \rangle / \langle n \rangle $]. The
suppression of its elements giving the probabilities of odd photon
numbers represents its most striking feature \cite{Waks2004}. If
the field were composed only of photon pairs, these elements would
have been zero. However, the presence of noise photons conceals
this feature and so we can observe it only for the data set (a)
obtained under low illumination (see Fig.~\ref{fig6}b).
\begin{figure}      
 \vbox{\noindent (a)
 \resizebox{0.8\hsize}{!}{\includegraphics{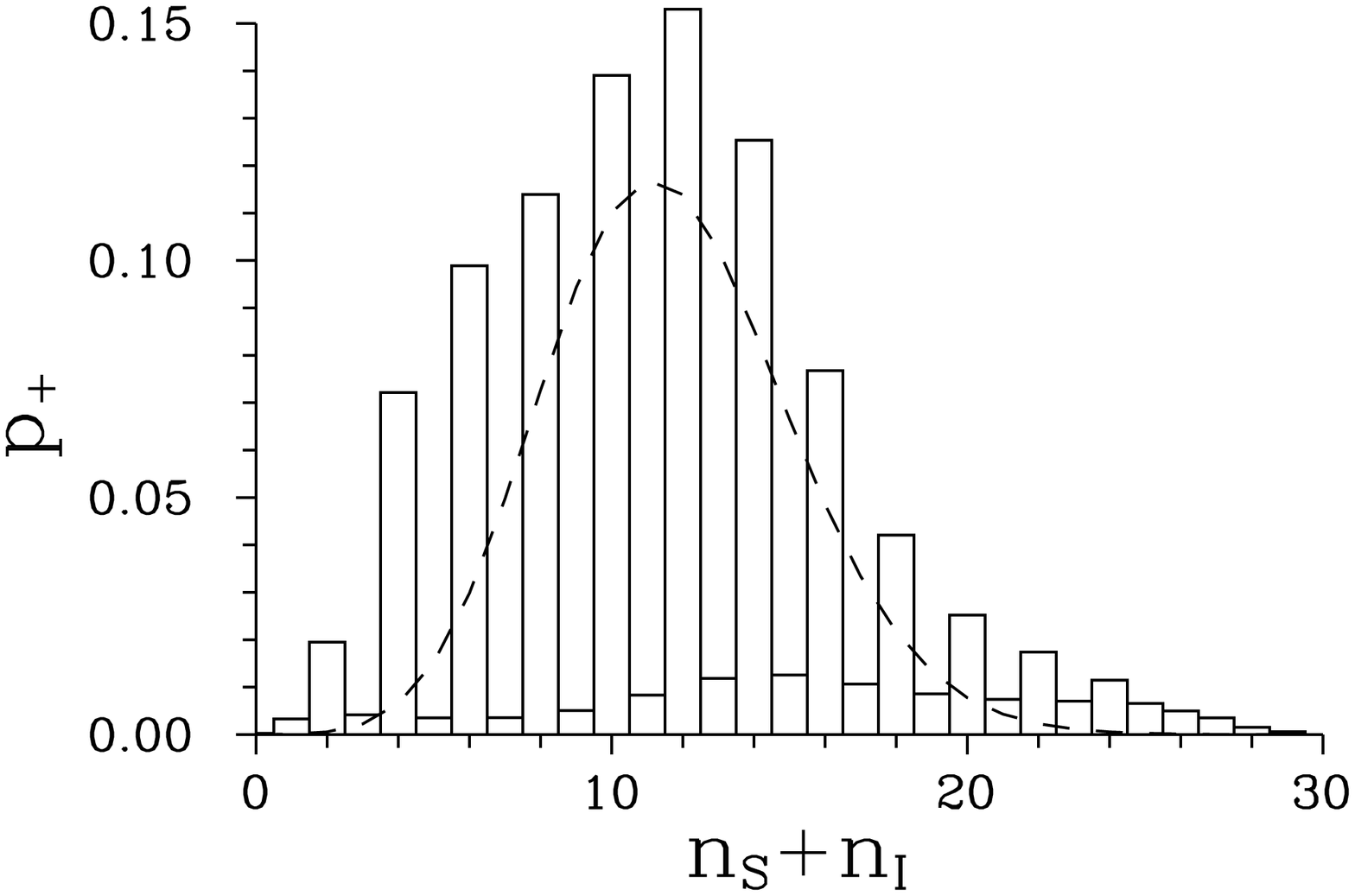}}
  \vspace{6mm}

 (b)
 \resizebox{0.8\hsize}{!}{\includegraphics{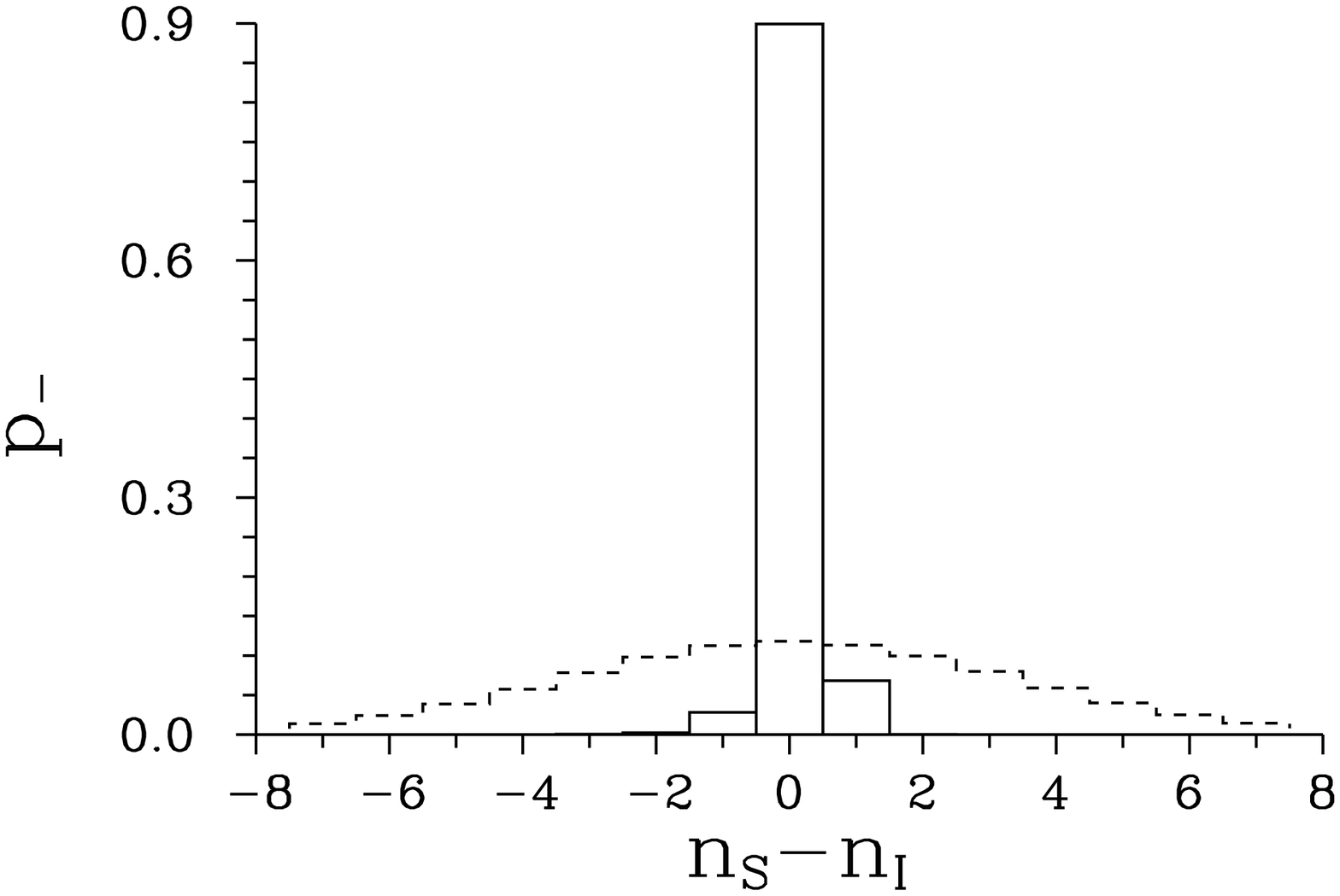}}
   }
 \vspace{3mm}
 \caption{a) Distributions $ p_{-} $ of the photon-number
  difference $ n_S - n_I $ and b) distributions $ p_{+} $ of the photon-number
  sum $ n_S + n_I $ for data set (b) (solid curves). Dashed curves
  give the distributions derived from the joint signal-idler photon-number
  distribution given as a direct product of the marginal signal
  and idler distributions; they are plotted for comparison.}
\label{fig6}
\end{figure}

\subsection{Determination of photon-number statistics, measurement of intense
fields, and role of the intensity profile}

The type of statistics of the emitted photon pairs is an important
characteristic \cite{Perina1991,Mandel1995}. According to the
theory, if the emission occurs in one spatiotemporal mode, the
photon-number statistics is Gaussian (thermal). However, the emission is usually observed
into more than one independent spatiotemporal modes
\cite{Avenhaus2008} and, as a consequence, the statistics of
photon pairs declines towards a Poissonian distribution. The
greater the number of modes, the closer the actual statistics to
the Poissonian distribution. In the experiment the situation is
more complex because of noises superimposed on the emitted paired
field. The theory presented in Sec.~VI below allows in principle
to determine the number of paired modes and so extract the
statistics of photon pairs. When the reconstruction algorithm is
applied, we can only determine the statistics of marginal signal
and idler fields and deduce the type of statistics of photon pairs
from them. We note that a Fano factor $ F $ is commonly used to
judge the type of statistics, or more precisely, the declination
of statistics from the Poissonian one.

The Fano factor is also extraordinarily important for the
quantification of the effect of presence of more than one photon
in the area of one macro-pixel at the photocathode. If the
probability of having more than one photon at one macro-pixel is
non-negligible the statistics of photoelectron numbers [$
f(c_S,c_I) $] decline from photon-number statistics [$ p_{\rm
rec}(n_S,n_I) $]. The fact that one macro-pixel cannot resolve
photon numbers leads to a systematic decrease of the Fano factor
of a detected field. The stronger the field the smaller the value
of the Fano factor. This effect is in its nature the same as a
dead-time effect in time-resolved detection. However, this effect
can be corrected using appropriate transfer matrices. When
stronger fields are measured, the transfer matrices $ K^{S} $ and
$ K^{I} $ should even be corrected with respect to the field
intensity profile as suggested in Sec.~IV.

Data set (c) has been obtained in the discussed regime and the
regions-of-interest were composed of $ N_S = N_I = 6528 $
macro-pixels. Here, the Fano factors of the marginal distributions
of detected photoelectrons are $ F_S = 0.996 $ and $ F_I = 1.008
$. We can see in Fig.~\ref{fig7} how the Fano factors $ F_S $ and
$ F_I $ of the marginal distributions derived from the
reconstructed joint signal-idler photon-number distribution $
p_{\rm rec} $ depend on the form of transformation matrices $
K^{S,N_S} $ and $ K^{I,N_I} $, in more detail on parameter $ M $
($ M = M_S = M_I $) giving the number of areas inside the
detection region. In the $ k $-th area there are pixels
illuminated by intensities greater than $ (k-1)\Delta I $ and
lower than $ k\Delta I $, $ k=1,\ldots,M $ [$ \Delta I = I_{\rm
max} / M $, $ I_{\rm max} $ being the maximum intensity found in
the profile]. The Fano factors $ F_S $ and $ F_I $ plotted for $ M
= 0 $ in Fig.~\ref{fig7} are obtained assuming the transfer
matrices given in Eq.~(\ref{10}) [$ N_S, N_I \longrightarrow
\infty $].  We can see from the curves in Fig.~\ref{fig7} that the
more precise the form of transfer matrices the better the
elimination of the effect and so the greater the values of Fano
factors $ F_S $ and $ F_I $. We can also deduce that it is
sufficient to divide the detection region-of-interest into several
areas to arrive at reliable results. The reconstruction of fields
described by data set (c) clearly demonstrates the ability of the
method to cope with this problem.

The Fano factors of marginal distributions have been determined as
$ F_S = 1.32 $ (a), 1.126 (b), and 1.106 (c) and $ F_I = 1.33 $
(a), 1.126 (b), and 1.165 (c). These values show that the observed
down-conversion has been emitted into several tens or even
hundreds of independent spatiotemporal modes.
\begin{figure}      
 \resizebox{0.8\hsize}{!}{\includegraphics{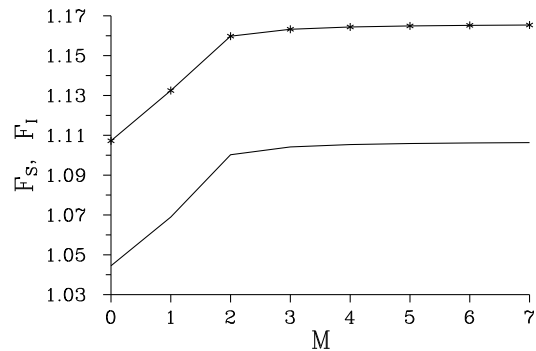}}
 \vspace{3mm}
 \caption{Fano factors $ F_S $ (solid curve) and $ F_I $ (solid curve with $ \ast $)
  of the marginal signal and idler photon-number distributions obtained from the
  reconstructed joint signal-idler photon-number distribution $ p_{\rm rec} $ as they depend on the
  number $ M $ of areas defined inside the detection region-of-interest. }
\label{fig7}
\end{figure}

Provided that the generated joint signal-idler field is multi-mode
and the photons are preferably generated in the spontaneous
process, we can assume that relative phases of different modes
have random values. In this case, the joint signal-idler field can
fully be characterized by a joint signal-idler quasi-distribution
of signal and idler integrated intensities $ P_W $. The
quasi-distribution $ P_W $ can be uniquely derived from the joint
signal-idler photon-number distribution $ p_{\rm rec} $ using the
decomposition into Laguerre polynomials
\cite{Perina1991,Haderka2005a}. Due to pairwise character of the
emitted fields, the joint signal-idler quasi-distributions $ P_W $
of integrated intensities attain negative values in certain
regions \cite{Perina2009,Perina2011}.

\section{Fit of the experimental data using the model of signal
and noise}

An alternative approach for obtaining a joint signal-idler
photon-number distribution in the output plane of the crystal can
be developed assuming a certain form of this distribution. We can
assume for physical reasons that the overall field can be
described by a certain form of superposition of signal and noise
and can also be decomposed into three independent contributions.
The first contribution describes photon pairs that are inside $
m_p $ independent modes with mean photon-pair numbers $ b_p $. The
second (third) contribution takes into account the presence of
noise in the signal (idler) field and is composed of $ m_S $ ($
m_I $) independent modes with mean photon numbers $ b_S $ ($ b_I
$).

On the experimental side of the problem there are five first- and
second-order moments of the measured photoelectron numbers: $
\langle c_S \rangle $, $ \langle c_I \rangle $, $ \langle c_S^2
\rangle $, $ \langle c_I^2 \rangle $, and $ \langle c_S c_I
\rangle $. Moreover a reliable determination of overall detection
efficiencies $ T_S\eta_S $ and $ T_I\eta_I $ in the experimental
setup is difficult. That is why we can consider the efficiencies $
T_S\eta_S $ and $ T_I\eta_I $ as parameters that should be
determined from the experimental data. We thus have eight
independent parameters and five measured quantities. The
requirement of minimum entropy of the joint photoelectron
distribution used for fitting the experimental data can be applied
to find the most appropriate form of the joint signal-idler
photon-number distribution $ p $ (details can be found in a
forthcoming publication).

The application of the method to data set (b) has provided the
values $ T_S\eta_S = 0.207 $ and $ T_I\eta_I = 0.205 $ that were
used in the reconstruction in Sec.~V. Values of the remaining
parameters have been determined as $ m_p = 628 $, $ b_p = 0.066 $,
$ m_S = 0.46 $, $ b_S = 0.173 $, $ m_I = 0.018 $, and $ b_I = 2.32
$. Comparison of the obtained joint signal-idler photon-number
distribution $ p_{\rm fit} $ with the distribution $ p_{\rm rec} $
revealed by the reconstruction is given in Fig.~\ref{fig8}. We can
see that the distribution $ p_{\rm fit} $ is 'narrower', i.e. it
contains less noise ($ F_S = 1.066 $, $ F_I = 1.068 $). This is in
agreement with the expectation that the fitting method is by its
nature more efficient in eliminating the noise. We note for
comparison that using the fitting method covariance $ C $ of the
signal and idler photon numbers $ n_S $ and $ n_I $ equals 99.7~\%
and $ R = 0.028 $. However, compared to the reconstruction method,
the fitting method does not take into account the number of
macro-pixels and also cannot be applied to more intense fields.
\begin{figure}      
 \vbox{\noindent (a)
 \resizebox{0.8\hsize}{!}{\includegraphics{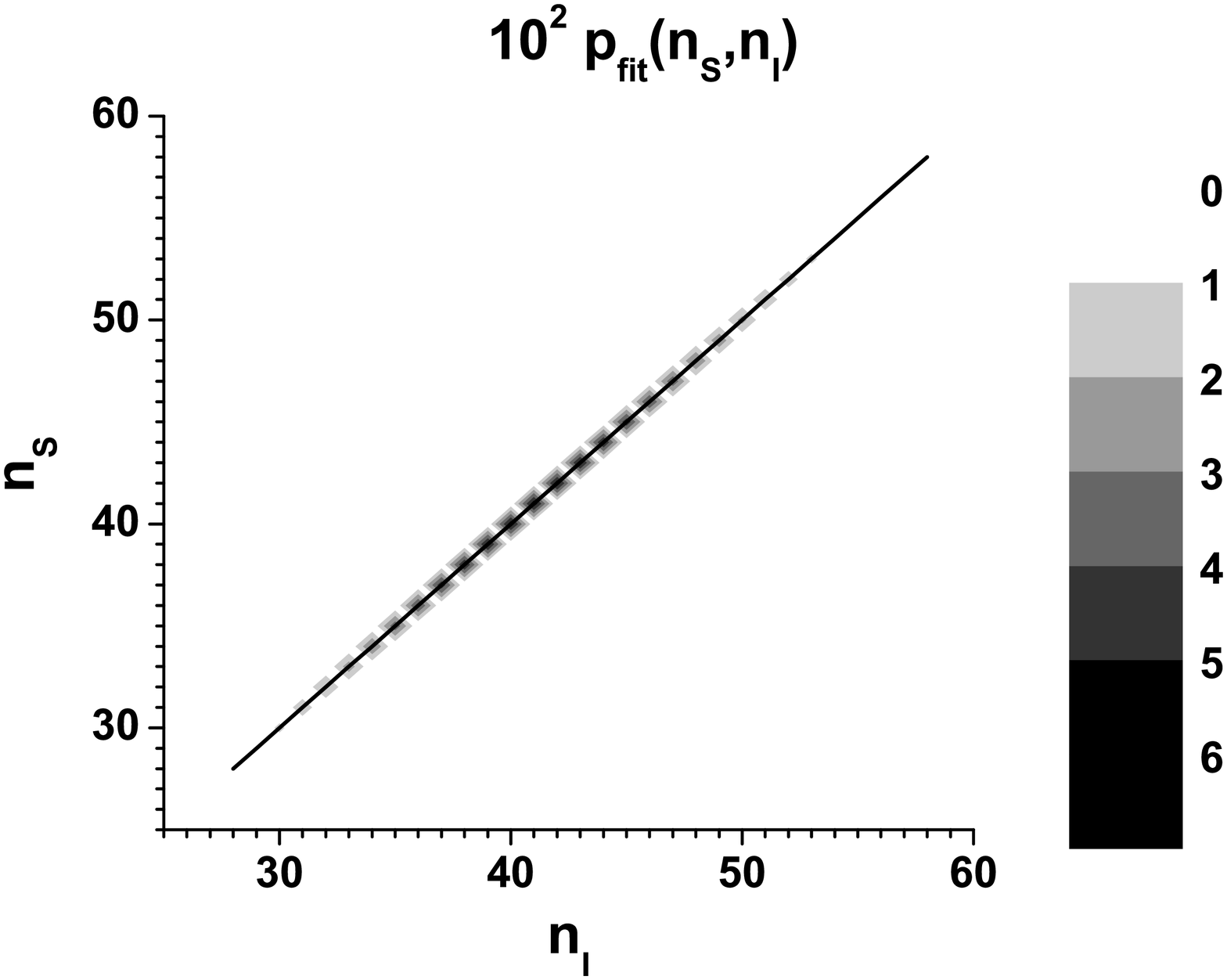}}
  \vspace{3mm}

 (b)
 \resizebox{0.8\hsize}{!}{\includegraphics{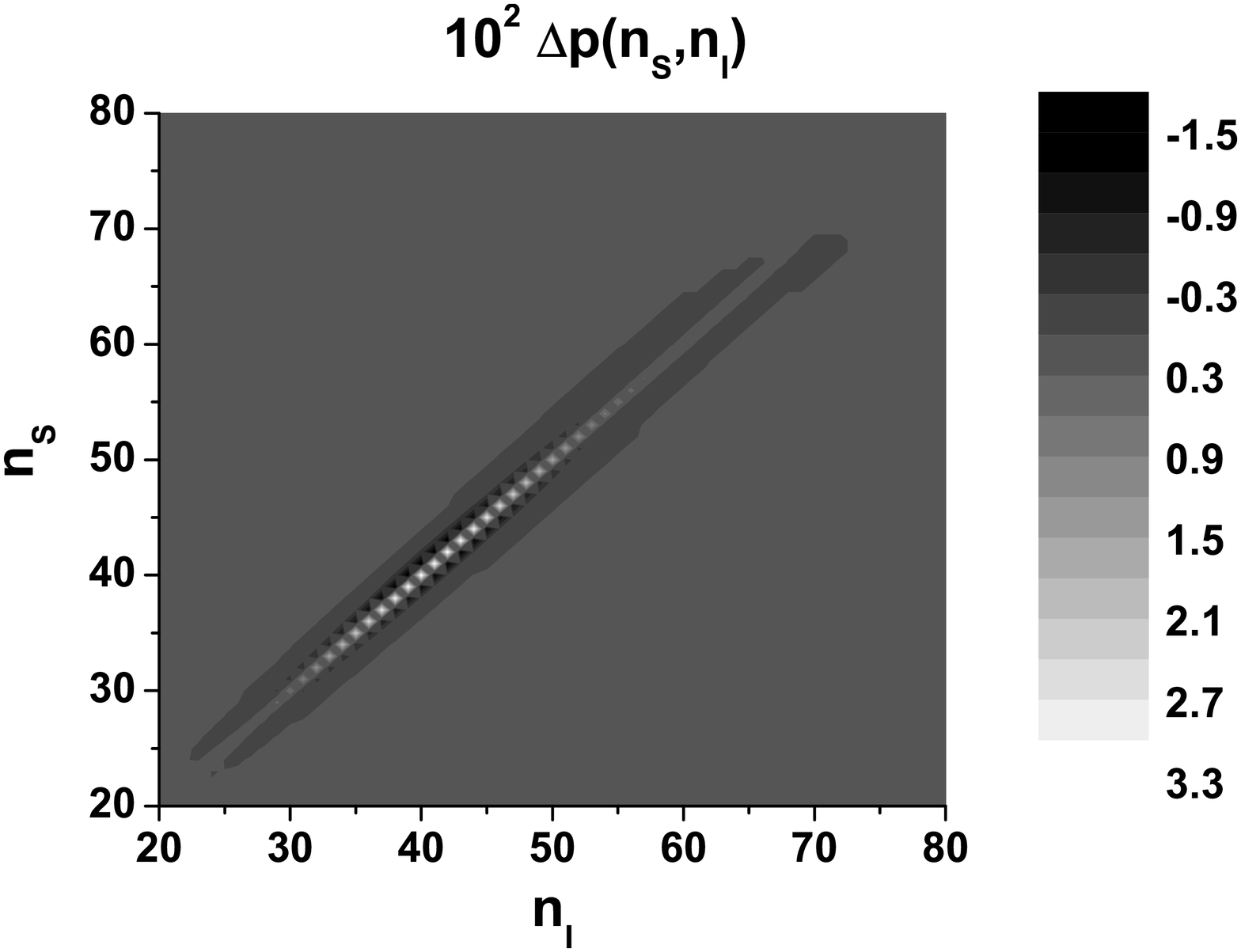}}
   }
 \vspace{3mm}
 \caption{Topological graphs of a) joint signal-idler photon-number distribution
  $ p_{\rm fit}(n_S,n_I) $ obtained by the fitting method and b) difference $ \Delta p(n_S,n_I) $
  of photon-number distributions $ p_{\rm fit} $ and $ p_{\rm rec} $ obtained by the
  fitting and reconstruction methods, respectively
  [$ \Delta p(N_S,N_I) = p_{\rm fit}(n_S,n_I) - p_{\rm rec}(n_S,n_I) $] for data set (b).
  The black line identifies the elements $ p_{\rm fit}(n_S,n_I) $ violating the
  classical inequality (\ref{32}).}
\label{fig8}
\end{figure}

\section{Conclusions}

We have developed a method for the reconstruction of a joint
signal-idler photon-number distribution using the measured
histograms of photoelectron numbers and an iteration expectation
maximization algorithm. In the framework of a general detection
theory we have found formulas for the transfer matrices that give
linear relations between elements of a photon-number distribution
and the corresponding photoelectron distribution. These formulas
take into account finite quantum detection efficiencies, the level
of dark counts as well as finite numbers of detection
macro-pixels. Special formulas appropriate for very weak as well
as high illumination intensities have been found. A method for the
inclusion of a transverse intensity profile into the form of
transfer matrices has been suggested. Three joint signal-idler
photon-number distributions differing in mean photon-numbers have
been reconstructed using the developed method. Some of their
elements violate a classical inequality. Fluctuations of the
difference of signal and idler photon numbers are highly
suppressed due to pairing of photons in all three cases
(sub-shot-noise correlations). Moreover there occurs a partial
suppression of elements corresponding to odd photon numbers in the
distribution of the sum of signal and idler photon numbers for the
weakest measured field. The developed reconstruction method has
been compared to a method that provides the best fit of the
experimental data assuming a joint signal-idler photon-number
distribution in the form of superposition of signal and noise. The
power of the reconstruction method to eliminate noise has been
found weaker on one side. On the other side, it allows more
realistic description of the detection process. This is invaluable
for higher detector illumination intensities. We believe that the
developed reconstruction method will stimulate a broader use of
iCCD cameras as photon-number-resolving detectors.

\acknowledgments The authors thank J. Pe\v{r}ina for helpful
discussions. Support by projects 1M06002, COST OC 09026 and
Operational Program Research and Development for Innovations -
European Regional Development Fund (project CZ.1.05/2.1.00/03.0058) of the
Ministry of Education of the Czech Republic as well as project
IAA100100713 of GA AV \v{C}R is acknowledged.

\appendix

\section{Determination of an effective detection quantum efficiency}

We assume that a Poissonian field with mean photon number $ \mu $
and statistical operator $ \hat{\varrho} $,
\begin{equation}   
 \hat{\varrho} = \sum_{n=0}^{\infty} \frac{\mu^n}{n!} \exp(-\mu)
 |n\rangle \langle n| ,
\end{equation}
impinges on a detector with quantum efficiency $ \eta $ and
dark-count rate $ d $. The probability $ p^{\rm Pois} $ of
registering a photon is given as
\begin{eqnarray}  
 p^{\rm Pois} &=& {\rm Tr}(\hat{D}\hat{\varrho}) \nonumber \\
 &=&  1-(1-d)\exp(-\eta\mu) ;
\label{A2}
\end{eqnarray}
the detection operator $ \hat{D} $ has been introduced in
Eq.~(\ref{3}).

If there is just one photon in the Fock state ($ \hat{\varrho} =
|1\rangle \langle 1| $) in a detected field, the probability $
p^{\rm Fock} $ of its counting equals
\begin{equation}   
 p^{\rm Fock} = 1-(1-d)(1-\eta) .
\end{equation}
The requirement of equal detection probabilities $ p^{\rm Fock} $
and $ p^{\rm Pois} $ results in an effective quantum efficiency $
\eta^{\rm eff} $ depending on $ \mu $:
\begin{equation}   
 \eta^{\rm eff}(\mu) = 1-\exp(-\eta\mu) .
\label{A4}
\end{equation}

\bibliography{perina}
\bibliographystyle{apsrev}

\end{document}